\documentclass{article}
\usepackage{amssymb}
\usepackage{amsmath}

\usepackage{epsf}
\usepackage{epsfig}
\usepackage{floatflt}
\usepackage{float}

\newcounter{sect} \numberwithin{sect}{section}
 \numberwithin{subsect}{subsection}

\newtheorem{theorem}[sect]{Theorem}

\newtheorem{remark}[sect]{Remark}

\newcommand{\rank}{\operatorname{rank}}

\begin{document}

\begin{center}

\bigskip

{\Large 
\textbf{Time-dependent polynomials with one multiple root and new
solvable dynamical systems}}

\bigskip

{\large Oksana Bihun\footnote{obihun@uccs.edu} \\
\normalsize University of Colorado, Colorado Springs, USA}
\bigskip

\begin{abstract}

A time-dependent monic polynomial  in the $z$ variable with $N$ distinct roots such that  exactly one  root has multiplicity $\bar{m} \geq 2$ is considered. For $k=1,2$, the $k$-th derivatives of the $N$  roots  are expressed in terms of the derivatives of order $j\leq k$ of the first $N$ coefficients of the polynomial and of the derivatives of order $j\leq k-1$ of the roots themselves. These relations are utilized to construct new classes of algebraically solvable first  order systems of ODEs as well as $N$-body problems. Multiple examples of solvable isochronous (all solutions are periodic with the same period) $2$- and $3$-body problems are provided.

\end{abstract}

\end{center}



\section{Introduction and main results}

In the last several decades, solvability or integrability of multiple dynamical systems has been proven~(see, for example, \cite{1,2,28} and references therein), the Calogero-Moser~\cite{3,4}, Sutherland~\cite{5,6} and goldfish~\cite{7} models among them.
Many of these systems have been constructed by exploiting the relation between the zeros and the coefficients of a monic time-dependent polynomial with distinct and simple roots. The main idea of this approach is that a solvable evolution of the coefficients of the polynomial must yield solvable evolution of its roots. 

In many cases~~\cite{1,2,8,9}, this main idea was implemented via  construction of a linear partial differential equation (PDE) that possesses a time-dependent polynomial solution. Such a PDE governs the evolution of both the coefficients and the zeros of its polynomial solution. Due to the linearity of the PDE, the evolution of the coefficients is then described by a linear, thus solvable, system of ordinary differential equations (ODEs). But then the system that describes the nonlinear evolution of the zeros of the polynomial  is algebraically solvable: Its solutions can be recovered via the algebraic operation of finding the roots of a monic polynomial with time-dependent coefficients that themselves can be obtained via algebraic operations. One of the technical challenges in this process is to obtain, in explicit form, the nonlinear system that governs the evolution of the zeros of the time-dependent polynomial solution of the PDE.

Recently, new formulas that explicitly express the $k$-th derivatives of the simple zeros of a monic time-dependent polynomial in terms of the $k$-th derivatives of its coefficients and the derivatives of order $j\leq k-1$ of the zeros themselves have been discovered~\cite{10,11}. In particular, for the $t$-dependent monic polynomial in the $z$ variable
\begin{eqnarray}
 &&q_{N}(z;t)=\prod_{n=1}^N (z-x_n)
=z^N+\sum_{n=1}^N \xi_n \,z^{N-n}\label{Polyxi}
\end{eqnarray}
with $N$ simple zeros $x_n=x_n(t)$ and $N$ coefficients $\xi_n=\xi_n(t)$, the formulas read~\cite{10}
\begin{eqnarray}
&&\dot{x}_n=-\left[\prod_{\ell=1, \ell \neq n}^N (x_n-x_\ell)^{-1} \right]
\sum_{j=1}^{N} (x_n)^{N-j} \dot{\xi}_j , \label{xnxidot}\\
&&\ddot{x}_n=\sum_{\ell=1, \ell \neq n}^N \frac{2 \dot{x}_n \dot{x}_\ell}{x_n-x_\ell} -
\left[\prod_{\ell=1, \ell \neq n}^N (x_n-x_\ell)^{-1} \right]
\sum_{j=1}^N (x_n)^{N-j} \ddot{\xi}_j, \label{xnxiddot}
\\
&& 1\leq n\leq N.\notag
\end{eqnarray}
The discovery of these formulas made it possible to take a nonlinear (compared to linear in the previous technique) solvable evolution of the coefficients $\xi_n$ as a point of departure in the construction of another nonlinear system  that governs the evolution of the zeros $x_n$.
This technique has been utilized in construction of new solvable evolution equations, including ordinary and partial differential as well as difference equations~\cite{12,13,14, 20,16,17,18,22}. In fact, the technique allows to construct infinite hierarchies of solvable systems of nonlinear evolution equations~\cite{14,20}, a remarkable find given that integrable systems are rare.

Even more recently, formulas~\eqref{xnxidot} and~\eqref{xnxiddot} have been generalized to the case where polynomial~\eqref{Polyxi} has exactly one root of multiplicity two~\cite{23}.  Using this generalization, new classes of solvable first and second order nonlinear systems of ODEs have been constructed. However, the method that was used to obtain this particular generalization of relations~\eqref{xnxidot} and~\eqref{xnxiddot} does not seem to have an obvious extension to the case where  polynomial~\eqref{Polyxi} has a root of multiplicity higher than two.

In this paper, the case of  a time-dependent monic polynomial in the $z$ variable with exactly one root of multiplicity $\bar{m}\geq 2$ is considered. A method different from that of~\cite{23} is utilized to derive generalizations of formulas~\eqref{xnxidot} and~\eqref{xnxiddot} for this case. New classes of solvable first and second order systems of nonlinear evolution equations are constructed. For the case where $\bar{m}=2$, these systems are equivalent to some of the systems reported in~\cite{23}, see Remarks~\ref{rem:m11order1_N},~\ref{rem:m11_N} and~\ref{rem:m11_2}. Several examples of solvable $2$- and $3$-body problems are provided, all of them (possibly asymptotically) isochronous~\cite{24}, that is, such that all their solutions are (possibly asymptotically as $t \to \infty$) periodic with the same period independent of the initial data. Solutions of these $N$-body problems are plotted.

Let $N \geq 2$ and $\bar{m} \geq 2$ be fixed  integers, let $m_1= \bar{m}-1$ and let $t$ be the ``time'' variable. Consider the following time-dependent monic polynomial of degree $N+\bar{m}-1=N+m_1$ in the complex variable~$z$:
\begin{subequations}
\begin{eqnarray}
 &&p_{N+m_1}(z;t)=(z-x_1)^{m_1}\prod_{n=1}^N (z-x_n)\label{MainPolyx} \\
 &&=(z-x_1)^{m_1}(z^N+\sum_{n=1}^N \xi_n \,z^{N-n})\label{MainPolyXim1}\\
&&  =z^{N+m_1}+\sum_{n=1}^{N+m_1} y_n\, z^{N+m_1-n},\label{MainPolyy}
\end{eqnarray}
\label{MainPoly}
\end{subequations}
where $x_n=x_n(t)$, $\xi_n=\xi_n(t)$ for $n=1,2,\ldots, N$,  and 
$y_n=y_n(t)$ for $n=1,2,\ldots, N+m_1$. Assume that the $N$ roots $x_1, \ldots, x_N$ of this polynomial are distinct for all $t$, while noting the fact that the root $x_1=x_1(t)$ has multiplicity $\bar{m}=m_1+1$.
 


In this paper,  the $k$-th time-derivatives, for $k=1,2$, of the zeros $\{x_n\}_{n=1}^N$ of polynomial~\eqref{MainPoly}  are expressed explicitly in terms of  the derivatives of order $j \leq k$ of the first $N$ coefficients $\{y_n\}_{n=1}^N$ of the polynomial and of the derivatives of order $ j \leq k-1$ of the zeros themselves, see~\eqref{xndotFormula} and~\eqref{xnddotFormula} in Theorems~\ref{Theorem1stOrder},~~\ref{Theorem2ndOrder}.
The significance of the detail that the formulas involve only the first $N$ coefficients of polynomial~\eqref{MainPoly}  stems from the following observation: Because the $N+m_1$ coefficients $\{y_n\}_{n=1}^{N+m_1}$ can be expressed in terms of the $N$ zeros $\{x_n\}_{n=1}^N$ via the Vieta relations, at most $N$ among these coefficients are functionally independent. This means that the evolution of $N$ among the coefficients $\{y_n\}_{n=1}^{N+m_1}$ determines the evolution of the remaining $m_1$ coefficients. Therefore, while in this setting we expect solvable evolution of the coefficients of polynomial~\eqref{MainPoly} to yield solvable evolution of its zeros, the evolution of the coefficients $\{y_n\}_{n=1}^{N+m_1}$ cannot be assigned freely, but rather, the evolution of only $N$ among these coefficients is to determine the evolution of the remaining $m_1$ coefficients and therefore the evolution of the zeros $\{x_n\}_{n=1}^N$.

The generalizations of formulas~\eqref{xnxidot} and~\eqref{xnxiddot} described in the previous paragraph are  used to construct solvable first and second order nonlinear systems of ODEs, which are presented, again, in Theorems~\ref{Theorem1stOrder},~\ref{Theorem2ndOrder}.  Using Theorem~\ref{Theorem2ndOrder}, several solvable $2$- and $3$-body problems are constructed, all of them (possibly asymptotically) isochronous. 

To formulate the main results, let us introduce some additional notation. For $1 \leq n, m \leq N$, let 
\begin{subequations}
\begin{eqnarray}
\alpha_{nm}=(-1)^{n+m+1} \sum_{k=1}^{n-m} \beta_{nm}^{(k)},
\label{alphanm}
\end{eqnarray}
where the quantities $\beta_{nm}^{(k)}$ are defined recursively as follows:
\begin{eqnarray}
&&\beta_{nm}^{(k)}=- \sum_{j=m+1}^{n+1-k} \binom{m_1}{j-m} \beta_{nj}^{(k-1)}, \; k=2,3,\ldots,\notag\\
&&\beta_{nm}^{(1)}=\left\{ 
\begin{array}{l}
\binom{m_1}{n-m} \mbox{ if } m\leq n-1,\\
0 \mbox{ if } m\geq n
\end{array}
 \right.
\label{betanm}
\end{eqnarray}
\label{alphabetanm}
\end{subequations}
and $\binom{n}{m}$ denotes the binomial coefficient, which is assumed to vanish if $m>n$ or $m<0$.

Let us also define 
\begin{eqnarray}
&&\gamma_n=
\sum_{j=1}^{n-1}  (-1)^j \binom{m_1}{j} \alpha_{nj} +(-1)^n \binom{m_1}{n},\notag\\
&&1\leq n\leq N.
\label{gamman}
\end{eqnarray}

Let $\sigma_n^{(m)}$ be the symmetric polynomial of degree $n$ in $m$ variables defined by
\begin{eqnarray}
\sigma_n^{(m)}(\eta_1, \ldots, \eta_m)=\sum_
{\begin{array}{c}
j_1+\cdots+j_m=n\\ j_\nu \in \{0,1\}
\end{array}}
(\eta_1){}^{j_1}\cdots (\eta_m){}^{j_m},
\label{sigmanm}
\end{eqnarray}
where the sum is taken over all the $m$-tuples $(j_1, \ldots, j_m)$ of indices  $j_\nu$ having values $0$ or $1$ and satisfying $j_1+\cdots+j_m=n$. For convenience, assume that $\sigma^{(m)}_0=1$ and that $\sigma^{(m)}_n=0$ if $n>m$.

The main results of the paper are stated in the following two theorems.

\begin{theorem} 
For $\vec{x}=(x_1, \ldots, x_N)$,  $\vec{y}^{(N)}=(y_1, \ldots, y_N)$,  $\dot{\vec{y}}^{(N)}=(\dot{y}_1, \ldots, \dot{y}_N)$, $\underline{\alpha}$  the $N\times N$ matrix with the components $\alpha_{mj}$  defined by~\eqref{alphabetanm} and  $\vec{\gamma}=(\gamma_1, \ldots, \gamma_N)$ defined by~\eqref{gamman},
let
\begin{eqnarray}
\label{hn1}
&&h_1^{(1)}(\vec{x}; \vec{y}^{(N)}, \dot{\vec{y}}^{(N)})\equiv h_1^{(1)}(\vec{x}; \vec{y}^{(N)}; \dot{\vec{y}}^{(N)};\vec{\gamma}, \underline{\alpha})\notag\\
&&
=-\left[
(m_1+1)! \prod_{\ell=2}^N(x_1-x_\ell)
\right]^{-1}\sum_{j=1}^N (N-j+1)_{m_1} (x_1)^{N-j} \dot{y}_j
\end{eqnarray}
and let
\begin{eqnarray}
&&h_n^{(1)}(\vec{x}; \vec{y}^{(N)}; \dot{\vec{y}}^{(N)})\equiv h_n^{(1)}(\vec{x}; \vec{y}^{(N)}; \dot{\vec{y}}^{(N)}; \vec{\gamma}, \underline{\alpha})\notag\\
&&=-\left[\prod_{\ell=1, \ell \neq n}^N (x_n-x_\ell) \right]^{-1}
\sum_{m=1}^{N} \Bigg\{ (x_n)^{N-m} \Big[
\dot{y}_m+ \sum_{j=1}^{m-1} \alpha_{mj} (x_1)^{m-j} \dot{y}_j \Big]
\Bigg\}\notag\\
&&-(x_n-x_1)^{-2}\left[ (m_1+1)! \prod_{\ell=2, \ell \neq n}^N (x_n-x_\ell)(x_1-x_\ell)\right]^{-1}\notag\\
&& \cdot
\left[ \sum_{k=1}^N (N-k+1)_{m_1} (x_1)^{N-k}\dot{y}_k
\right]\notag\\
&& \cdot
\left\{
\sum_{m=1}^N (x_n)^{N-m}
\left[ -m \gamma_m (x_1)^{m-1} +\sum_{j=1}^{m-1} \alpha_{mj} (m-j) (x_1)^{m-j-1} y_j 
\right]
\right\},\notag\\
&&2\leq n\leq N. 
\end{eqnarray}
If $\vec{x}=\big(x_1, \ldots, x_N\big)$ is a vector of the zeros of polynomial~\eqref{MainPoly}, where the zero $x_1$ has multiplicity $\bar{m}$, then the first derivatives $\dot{x}_n$ of the zeros can be expressed explicitly in terms of the vector $\vec{y}^{(N)}=\big(y_1, \ldots, y_N \big)$ of the first $N$ coefficients of the same polynomial and its first derivative $\dot{\vec{y}}^{(N)}$, as well as the zeros $\vec{x}$ themselves, as follows:
\begin{eqnarray}
&&\dot{x}_n=h_n^{(1)}(\vec{x}; \vec{y}^{(N)}; \dot{\vec{y}}^{(N)}; \vec{\gamma}, \underline{\alpha}),\;\; 1\leq n\leq N.
 \label{xndotFormula}
\end{eqnarray}
Moreover, if the system of evolution equations
\begin{eqnarray}
&&\dot{y}_n=f_n^{(1)}(y_1, \ldots, y_N), \;1\leq n \leq N, 
\label{ydot_SystemThm}
\end{eqnarray}
is algebraically solvable, then the system of nonlinear evolution equations 
\begin{eqnarray}
&&\dot{x}_n=h_n^{(1)}(\vec{x}; \vec{y}^{(N)}; \vec{f}^{(1)}(\vec{y}^{(N)}); \vec{\gamma}, \underline{\alpha}),\;\; 1\leq n\leq N, 
\notag\\
&&y_j=(-1)^j \sigma_j^{(N+m_1)}(\underbrace{x_1, \ldots, x_1}_{m_1 \mbox{ times}}, x_1, \ldots, x_N),\; 1\leq j\leq N,
 \label{xndotSolvableExplicit}
\end{eqnarray}
where $\sigma_n^{(m)}$ is defined by~\eqref{sigmanm} and $\vec{f}^{(1)}(\vec{y}^{(N)})=\Big(f_1^{(1)}(\vec{y}^{(N)}),\ldots, f_N^{(1)}(\vec{y}^{(N)}) \Big)$, is solvable.
\label{Theorem1stOrder}
\end{theorem}
\begin{remark} \label{rem:m11order1_N} Note that in the special case where $m_1=1$, that is, if $x_1$ is the root of polynomial~\eqref{MainPoly} of multiplicity~2, system~\eqref{xndotSolvableExplicit} is equivalent to system~(31),~(27a) of~\cite{23} with $\bar{m}=N+1$ (in the notation of~\cite{23}) and $\dot{y}_m$ replaced by $f_m^{(1)}(\vec{y}^{(N)})$, $m=1,2,\ldots,N$.
\end{remark}
\begin{remark}
 Theorem~\ref{Theorem1stOrder} is  valid even if system~\eqref{ydot_SystemThm} is not autonomous, that is,  if the functions $f_j^{(1)}$ depend not only on  $\vec{y}^{(N)}$, but also on the  variable $t$ explicitly.
\end{remark}

\begin{theorem} 
For  $\vec{x}=(x_1, \ldots, x_N)$,  $\dot{\vec{x}}=(\dot{x}_1, \ldots, \dot{x}_N)$, $\vec{y}^{(N)}=(y_1, \ldots, y_N)$,  $\dot{\vec{y}}^{(N)}=(\dot{y}_1, \ldots, \dot{y}_N)$, $\ddot{\vec{y}}^{(N)}=(\ddot{y}_1, \ldots, \ddot{y}_N)$, $\underline{\alpha}$  the $N\times N$ matrix with the components $\alpha_{mj}$ defined by~\eqref{alphabetanm} and $\vec{\gamma}=(\gamma_1, \ldots, \gamma_N)$ defined by~\eqref{gamman},
let
\begin{subequations}
\label{hn2}
\begin{eqnarray}
&&h_1^{(2)}(\vec{x}; \dot{\vec{x}}; \vec{y}^{(N)}; \dot{\vec{y}}^{(N)};  \ddot{\vec{y}}^{(N)})\equiv h_1^{(2)}(\vec{x}; \dot{\vec{x}}; \vec{y}^{(N)}; \dot{\vec{y}}^{(N)};
 \ddot{\vec{y}}^{(N)}; \vec{\gamma}, \underline{\alpha})\notag\\
&&
=-\left[ (m_1+1)! \prod_{\ell=2}^N (x_1-x_\ell) \right]^{-1} \sum_{j=1}^N (N-j+1)_{m_1} (x_1)^{N-j} \ddot{y_j}\notag\\
&& +\dot{x}_1 \sum_{n=2}^N \frac{m_1\dot{x}_1 +2 \dot{x}_n}{x_1-x_n}
\end{eqnarray}
and let
\begin{eqnarray}
&&h_n^{(2)}(\vec{x}; \dot{\vec{x}}; \vec{y}^{(N)}; \dot{\vec{y}}^{(N)};  \ddot{\vec{y}}^{(N)})\equiv h_n^{(2)}(\vec{x}; \dot{\vec{x}}; \vec{y}^{(N)}; \dot{\vec{y}}^{(N)};  \ddot{\vec{y}}^{(N)}; \vec{\gamma}, \underline{\alpha})\notag\\
&&=\sum_{\ell=1, \ell \neq n}^N \frac{2 \dot{x}_n \dot{x}_\ell}{x_n-x_\ell} -\Big[ \prod_{\ell=1, \ell \neq n}^N (x_n-x_\ell) \Big]^{-1}\notag\\
&&\cdot \Bigg\{
\sum_{m=1}^N (x_n)^{N-m} \ddot{y}_m + \sum_{m=1}^N \sum_{j=1}^{m-1} \alpha_{mj} (x_n)^{N-m}(x_1)^{m-j} \ddot{y}_j\notag\\
&&-(\dot{x}_1)^2 \sum_{m=1}^N m(m-1) \gamma_m (x_n)^{N-m} (x_1)^{m-2}\notag\\
&&+(\dot{x}_1)^2 \sum_{m=1}^N \sum_{j=1}^{m-2} (m-j) (m-j-1) \alpha_{mj} (x_n)^{N-m} (x_1)^{m-j-2} y_j\notag\\
&&+2 \dot{x}_1 \sum_{m=1}^N \sum_{j=1}^{m-1} (m-j) \alpha_{mj} (x_n)^{N-m} (x_1)^{m-j-1}\dot{y}_j
\Bigg\}\notag\\
&&+\Bigg\{
(x_n-x_1)^{-2}
\Big[ 
(m_1+1)! \prod_{\ell=2, \ell \neq n} ^N (x_n-x_\ell)(x_1-x_\ell)
\Big]^{-1} \sum_{j=1}^N (N-j+1)_{m_1} (x_1)^{N-j} \ddot{y}_j\notag\\
&&+ \dot{x}_1 \Big[ \prod_{\ell=1, \ell \neq n}^N(x_n-x_\ell) \Big]^{-1} \sum_{j=2}^N \frac{m_1 \dot{x}_1+2 \dot{x}_j}{x_1-x_j}
\Bigg\}\notag\\
&&\cdot \Bigg\{
\sum_{m=1}^N m \gamma_m (x_n)^{N-m} (x_1)^{m-1}\notag\\
&&-\sum_{m=1}^N \sum_{j=1}^{m-1} (m-j) \alpha_{mj} (x_n)^{N-m} (x_1)^{m-j-1} y_j
\Bigg\},\notag\\
&&2\leq n\leq N.
\end{eqnarray}
\end{subequations}
If $\vec{x}=\big(x_1, \ldots, x_N \big)$ is a vector of the zeros of polynomial~\eqref{MainPoly}, where the zero $x_1$ has multiplicity $\bar{m}$, then the second derivatives $\ddot{x}_n$ of the zeros can be expressed explicitly in terms of the vector $\vec{y}^{(N)}=\big(y_1, \ldots, y_N \big)$ of the first $N$ coefficients of the same polynomial and its first and second derivatives $\dot{\vec{y}}^{(N)}$,  $\ddot{\vec{y}}^{(N)}$, as well as the zeros $\vec{x}$ themselves and their first derivatives $\dot{\vec{x}}$, as follows:
\begin{eqnarray}
\ddot{x}_n=h_n^{(2)}\Big(\vec{x}; \dot{\vec{x}}; \vec{y}^{(N)}; \dot{\vec{y}}^{(N)};  \ddot{\vec{y}}^{(N)}; \vec{\gamma}; \underline{\alpha}\Big),\;\; 1\leq n\leq N.
 \label{xnddotFormula}
\end{eqnarray}
Moreover, if  the system of evolution equations
\begin{eqnarray}
&&\ddot{y}_n=f_n^{(2)}(y_1, \ldots, y_N, \dot{y}_1, \ldots, \dot{y}_N), \;1\leq n \leq N,
\label{yddot_SystemThm}
\end{eqnarray}
is algebraically solvable, then the system of nonlinear evolution equations 
\begin{subequations}
\begin{eqnarray}
&&\ddot{x}_n=h_n^{(2)}\Big(\vec{x}; \dot{\vec{x}}; \vec{y}^{(N)}; \dot{\vec{y}}^{(N)}; \vec{f}^{(2)}(\vec{y}^{(N)}, \dot{\vec{y}}^{(N)}); \vec{\gamma}; \underline{\alpha}\Big),\;\; 1\leq n\leq N, \\
&&y_j=(-1)^j \sigma_j^{(N+m_1)}(\underbrace{x_1, \ldots, x_1}_{m_1 \mbox{ times}}, x_1, \ldots, x_N),\label{yjThm2}\\
&&
\dot{y}_j=(-1)^j \sum_{k=1}^N \left\{ \frac{\partial }{\partial x_k}\left[\sigma_j^{(N+m_1)}(\underbrace{x_1, \ldots, x_1}_{m_1 \mbox{ times}}, x_1, \ldots, x_N) \right] \dot{x}_k \right\}
\label{dotyjThm2}\\
&&1\leq j\leq N,\notag
\end{eqnarray}
 \label{xnddotSolvableExplicit}
\end{subequations}
where $\sigma_n^{(m)}$ is defined by~\eqref{sigmanm} and $\vec{f}^{(2)}(\vec{y}^{(N)},\dot{\vec{y}}^{(N)})=\Big(f_1^{(2)}(\vec{y}^{(N)},\dot{\vec{y}}^{(N)}),\ldots, f_N^{(2)}(\vec{y}^{(N)},\dot{\vec{y}}^{(N)}) \Big)$, is solvable.
\label{Theorem2ndOrder}
\end{theorem}
\begin{remark} \label{rem:m11_N} Note that in the special case where $m_1=1$, that is, if $x_1$ is the root of polynomial~\eqref{MainPoly} of multiplicity~2, $N$-body problem~\eqref{xnddotSolvableExplicit} is equivalent to system~(32) of~\cite{23} with $\bar{m}=N+1$ (in the notation of~\cite{23}) and $\ddot{y}_j$ replaced by $f_j^{(2)}(\vec{y}^{(N)}, \dot{\vec{y}}^{(N)})$, $j=1,2,\ldots,N$, see also Remark~\ref{rem:m11_2}.
\end{remark}
\begin{remark}
 Theorem~\ref{Theorem2ndOrder} is  valid even if system~\eqref{yddot_SystemThm} is not autonomous, that is,  if the functions $f_j^{(2)}$ depend not only on $\vec{y}^{(N)}, \dot{\vec{y}}^{(N)}$, but also on the  variable $t$ explicitly.
\end{remark}

The outline of the rest of the paper is the following. Section~\ref{Sect2} is dedicated to the proofs of the main results. Section~\ref{Sect3} contains  examples of solvable $2$- and $3$-body problems, all of them either isochronous or asymptotically isochronous. These $2$- and $3$-body problems are illustrated by solution plots. Section~\ref{Sect4} is dedicated to discussion of the results; it also  outlines directions for future investigations.

\section{Proofs}
\label{Sect2}

Our first goal is to derive formulas that express the first two derivatives of the $t$-dependent zeros $\{x_n\}_{n=1}^N$ of  polynomial~\eqref{MainPoly} in terms of the first $N$ coefficients $\{y_n\}_{n=1}^N$ and their derivatives as well as lower order derivatives of the zeros themselves. Note that while polynomial~\eqref{MainPoly} has $N+m_1$ coefficients $\{y_n\}_{n=1}^{N+m_1}$, our aim is to eliminate the last $m_1$ coefficients from the formulas. Indeed, because the coefficients $\{y_n\}_{n=1}^{N+m_1}$ are expressed in terms of only $N$ distinct zeros  $\{x_n\}_{n=1}^N$ of polynomial~\eqref{MainPoly} via the Vieta relations, it is possible to express the last $m_1$ coefficients $\{y_{N+k}\}_{k=1}^{m_1}$ in terms of the first $N$ coefficients $\{y_n\}_{n=1}^N$ of polynomial~\eqref{MainPoly} and its multiple zero $x_1$, see~\eqref{eq:y_Npk}.

Let us begin with finding relations between the $t$-dependent coefficients $\{y_n\}_{n=1}^{N+m_1}$ and $\{\xi_n\}_{n=1}^N$ defined in (\ref{MainPoly}). 
Recall that $\sigma_n^{(m)}$ is the symmetric polynomial of degree $n$ in $m$ variables, see~\eqref{sigmanm}, and
observe that by the Vieta relations for polynomial~\eqref{MainPoly},
\begin{subequations}
\begin{eqnarray}
\xi_n&=&(-1)^n \sigma_n^{(N)}(x_1, \ldots, x_N), \; 1\leq n\leq N,\\
y_n&=&(-1)^n \sigma_{n}^{N+m_1}(\underbrace{x_1, \ldots, x_1}_{m_1 \mbox{ times}}, x_1, \ldots, x_N)\\
&=&(-1)^n \sigma_{n}^{N-1+\bar{m}}(\underbrace{x_1, \ldots, x_1}_{\bar{m} \mbox{ times}}, x_2, \ldots, x_N), \; 1\leq n \leq N+m_1.
\end{eqnarray}
\label{XiEtaY}
\end{subequations}

We shall use the last observation  to express $\{y_n\}_{n=1}^{N+m_1}$ in terms of $\{\xi_n\}_{n=1}^N$ and $x_1$. To this end,  let us express symmetric polynomials with a repeated argument in terms of symmetric polynomials of lower degree:
\begin{eqnarray}
&&\sigma_{n}^{N+m_1}(\underbrace{x_1, \ldots, x_1}_{m_1 \mbox{ times}}, x_1, \ldots, x_N)\notag\\
&=&
 \sum_
{\begin{array}{c}
j_1+\cdots+j_{m_1}\\
+k_1+\ldots+k_N=n\\ j_\nu, k_\nu \in \{0,1\}
\end{array}}
(x_1){}^{j_1+\cdots+j_{m_1}} (x_1)^{k_1}\cdots (x_N){}^{k_N}\notag\\
&=&\sum_{j=0}^n \binom{m_1}{j} (x_1)^j 
 \sum_
{\begin{array}{c}
k_1+\ldots+k_N=n-j\\  k_\nu \in \{0,1\}
\end{array}}
 (x_1)^{k_1}\cdots (x_N){}^{k_N}\notag\\
&=&\sum_{j=0}^n \binom{m_1}{n-j} (x_1)^{(n-j)} \sigma_j^{(N)}(x_1, \ldots, x_N),\notag\\
&&1
\leq n \leq N+m_1.
\label{s12}
\end{eqnarray}
Using~(\ref{XiEtaY}) and (\ref{s12}), we obtain the desired relation between the coefficients $\{y_n\}_{n=1}^{N+m_1}$ and $\{\xi_n\}_{n=1}^N$:
\begin{eqnarray}
y_n=\sum_{j=0}^{n} \binom{m_1} {n-j} (x_1)^{n-j}(-1)^{n+j} \xi_j,
\label{YXi}
\end{eqnarray}
where it is assumed that $\xi_0=1$, $\xi_n=0$ if $n>N$. Recall also that $\binom {m}{k}=0$ if $k>m$.

Using the vector notation $\vec{y}=(y_1, \ldots, y_{N+m_1})^T$ and $\vec{\xi}=(\xi_1, \ldots, \xi_N)^T$, we rewrite (\ref{YXi}) as follows:
\begin{eqnarray}
\vec{y}&=&A(x_1)\,\vec{\xi}+\vec{a}(x_1),\label{Syst_yXi}
\end{eqnarray}
where $A=A(x_1)$ is an $(N+m_1)\times N$ matrix given componentwise by
\begin{eqnarray}
&&A_{nj}=A_{nj}(x_1)=\left\{
\begin{array}{l}
1 \mbox{ if } j=n,\\
\binom{m_1}{n-j}(-1)^{n+j} (x_1)^{n-j} \mbox{ if } j<n,\\
0 \mbox{ if } j>n
 \end{array}
\right.
\label{matrixA}
\end{eqnarray}
and $\vec{a}=\vec{a}(x_1)=(a_1, \ldots a_{N+m_1})^T$ is the $(N+m_1)$-vector with the components
\begin{eqnarray}
a_n=(-1)^n \binom{m_1}{n} (x_1)^n, \;\;1\leq n\leq N+m_1
\end{eqnarray}
(note that $a_n=0$ if $n>m_1$).

Our next task is to express the coefficients $\{\xi_n\}_{n=1}^N$ in terms of only the first $N$ coefficients $\{y_n\}_{n=1}^N$ of polynomial~\eqref{MainPoly}. Observe that
the upper $N \times N$ block of the matrix $A$, which we denote by $A^{(N)}$, is lower triangular with the main diagonal being the $N$-vector $(1, \ldots, 1)$. Thus, $A^{(N)}$ is invertible and $\rank A=N$. While relations   (\ref{Syst_yXi}), if viewed as a set of equations for the unknown $\vec{\xi}$, constitute an overdetermined system, that system is nevertheless consistent and has a unique solution because of how $\vec{\xi}$ and $\vec{y}$ are defined in (\ref{MainPoly}). The upper $N \times N$ block $A^{(N)}$ of the matrix $A$ is nonsingular, hence the last $m_1$ equations in system (\ref{Syst_yXi}) can be considered as redundant and $\vec{\xi}$ can be found from the first $N$ equations of the system. 

It can be verified (see Appendix~\ref{App1}) that the components of the inverse matrix $\left[A^{(N)}\right]^{-1}$ are given by
\begin{eqnarray}
&&\left[A^{(N)}\right]^{-1}_{nm}=\left[A^{(N)}\right]^{-1}_{nm}(x_1)\notag\\
&&=
\left\{
\begin{array}{l}
\delta_{nm} +(1-\delta_{nm})\alpha_{nm} (x_1)^{n-m} \mbox{ if } m \leq n \\
0 \mbox{ if } m>n,
\end{array}
\right.
\label{ANinverse}
\end{eqnarray}
where $1 \leq n,m \leq N$, $\delta_{nm}$ is the Kronecker symbol and $\alpha_{nm}$ are given by~\eqref{alphabetanm}.

In this context, recall that $\binom{m}{\ell}=0$ if $\ell>m$ and that a sum over an empty set of indices equals zero. Note  that 
\begin{eqnarray}
&&\beta_{nm}^{(k)}=0 \mbox{ if } m>n-k,\\
&&\alpha_{nm}=0 \mbox{ if } m\geq n,
\end{eqnarray}
see Appendix~\ref{App1}, thus formula (\ref{ANinverse}) may also be written as $\left[A^{(N)}\right]^{-1}_{nm}= \delta_{nm} +\alpha_{nm} (x_1)^{n-m}$.  Clearly, the matrix $\left[A^{(N)}\right]^{-1}$ is lower triangular with all  its diagonal entries being equal to $1$.

We thus express the vector of coefficients $\vec{\xi}=(\xi_1, \ldots, \xi_N)$  in terms of the first $N$ coefficients $\vec{y}^{(N)}=(y_1, \ldots, y_N)$ of polynomial~\eqref{MainPoly} and of its multiple zero $x_1$:
\begin{eqnarray}
\vec{\xi}= \left[A^{(N)}(x_1)\right]^{-1}\left[\vec{y}^{(N)}-\vec{a}^{(N)}(x_1)\right],
\end{eqnarray}
where $\vec{v}^{(N)}$ denotes the vector that consists of the first $N$ components of a vector~$\vec{v}$. In components,
\begin{eqnarray}
&&\xi_n=y_n+\sum_{j=1}^{n-1} \alpha_{nj} (x_1)^{n-j} y_j -\gamma_n (x_1)^n,
\label{XiInTOy}
\end{eqnarray}
where $\alpha_{nm}$ are defined by~\eqref{alphanm},~\eqref{betanm} and $\gamma_n$ is defined by~\eqref{gamman}.

By plugging $\xi_n$ from (\ref{XiInTOy}) into the last $m_1$ equations in system (\ref{Syst_yXi}), one can express $y_{N+1}, \ldots, y_{N+m_1}$ in terms of $y_1, \ldots, y_N$ and $x_1$ as follows:
\begin{subequations}
\begin{eqnarray}
&&y_{N+k}=(-1)^k \binom{m_1}{k} (x_1)^k y_N \notag\\
&&+
\sum_{j=1}^{N-1}\left[ \theta_{kj} (x_1)^{N+k-j} y_j\right] + (x_1)^{N+k} \phi_{k},\notag\\
&&1\leq k \leq m_1,
\end{eqnarray}
where
\begin{eqnarray}
&&\theta_{kj}=(-1)^{N+k+j} \binom{m_1}{N+k-j} +\sum_{\ell=j+1}^{N} \binom{m_1}{N+k-\ell} (-1)^{N+k+\ell} \alpha_{\ell j},\\
&&\phi_{k}=(-1)^{N+k}\Bigg\{\binom{m_1}{N+k}\notag\\
&&-\sum_{\ell=1}^N
\binom{m_1}{N+k-\ell} 
\Bigg[
\binom{m_1}{\ell} +\sum_{j=1}^{\ell-1} \binom{m_1}{j} (-1)^{\ell+j} \alpha_{\ell j}
\Bigg]
\Bigg\},\\
&&1\leq k\leq m_1, 1\leq j \leq N-1\notag
\end{eqnarray}
\label{eq:y_Npk}
\end{subequations}
and $\alpha_{n m}$ are defined by~\eqref{alphanm},~\eqref{betanm}.

A substitution of~\eqref{XiInTOy} into~\eqref{xnxidot} and~\eqref{xnxiddot} produces explicit formulas for $\dot{x}_n$  that contain $\dot{x}_1$ in the right-hand side and formulas for $\ddot{x}_n$ that contain $\ddot{x}_1$ in the right-hand side. Therefore, our next task is to express $\dot{x}_1$ in terms of $\{\dot{y}_n\}_{n=1}^N$, $\{x_n\}_{n=1}^N$ and $\ddot{x}_1$ in terms of $\{\ddot{y}_n\}_{n=1}^N$ , $\{x_n\}_{n=1}^N, \{\dot{x}_n\}_{n=1}^N$.

The application of the differential operators $ \frac{\partial^{m_1+1}}{\partial z^{m_1+1}}$,
$\frac{\partial}{\partial t} \frac{\partial^{m_1}}{\partial z^{m_1}}$ and $\frac{\partial^2}{\partial t^2} \frac{\partial^{m_1}}{\partial z^{m_1}}$ to the identity 
\begin{eqnarray}
(z-x_1)^{m_1}\prod_{n=1}^N (z-x_n) =z^{N+m_1}+\sum_{n=1}^{N+m_1} y_n z^{N+m_1-n},
\end{eqnarray}
see (\ref{MainPoly}), followed by the evaluation of the resulting identities at $z=x_1$, yields that $z=x_1$ is a simple root of the polynomial equation
\begin{eqnarray}
(N+1)_{m_1} z^N+\sum_{j=1}^N (N+1-j)_{m_1} y_j z^{N-j}=0
\label{PolyEqnx1}
\end{eqnarray}
and that the the first two time-derivatives of $x_1=x_1(t)$ are given by
\begin{eqnarray}
\dot{x}_1=-\left[
(m_1+1)! \prod_{n=2}^N(x_1-x_n)
\right]^{-1} \sum_{j=1}^N (N-j+1)_{m_1} (x_1)^{N-j} \dot{y}_j,
\label{x1dot}
\end{eqnarray}
and
\begin{eqnarray}
\ddot{x}_1=-\left[ (m_1+1)! \prod_{n=2}^N (x_1-x_n) \right]^{-1} \sum_{j=1}^N (N-j+1)_{m_1} (x_1)^{N-j} \ddot{y_j}\notag\\
 +\dot{x}_1 \sum_{n=2}^N \frac{m_1\dot{x}_1 +2 \dot{x}_n}{x_1-x_n},
 \label{x1ddot}
\end{eqnarray}
where $(\alpha)_m=\alpha (\alpha+1)\cdots(\alpha+m-1)$ is the Pochhammer symbol.
Therefore, $x_1, \dot{x}_1$ and $\ddot{x}_1$ are determined by $y_1, \ldots, y_N$. 

We may finally substitute relations~\eqref{XiInTOy} and~\eqref{x1dot},~\eqref{x1ddot} into~\eqref{xnxidot},~~\eqref{xnxiddot} to obtain the desired formulas for $\dot{x}_n$ and $\ddot{x}_n$. These formulas are listed in Theorems~\ref{Theorem1stOrder} and~\ref{Theorem2ndOrder}.

Our next task is to describe a method of construction of a solvable system of nonlinear ODEs for $x_1, \ldots, x_N$, while taking a solvable evolution of $y_1, \ldots, y_{N}$ as a point of departure. A {solvable first-order} nonlinear system of ODEs can be constructed as follows.
\begin{itemize}
\item[\textit{Step 1.}] Assign a {solvable} evolution of $y_1, \ldots, y_N$ via a system
\begin{eqnarray}
\label{System_ydot}
&&\dot{y}_n=f_n^{(1)}(y_1, \ldots, y_N), \; 1\leq n \leq N.\notag
\end{eqnarray}
\item[\textit{Step 2.}] Express $\dot{\xi}_1, \ldots, \dot{\xi}_N$ in terms of $y_1, \ldots, y_N$ and $x_1$ by using formulas (\ref{XiInTOy}) and substituting $f^{(1)}_n(y_1, \ldots, y_N)$ instead of $\dot{y}_n$, see (\ref{System_ydot}), and the right-hand side of (\ref{x1dot}) instead of $\dot{x}_1$, so that
\begin{eqnarray}
&&\dot{\xi}_n=g_n^{(1)}(y_1, \ldots, y_N, x_1), \label{xindot}\\
&&1\leq n \leq N.\notag
\end{eqnarray}
\item[\textit{Step 3.}] Substitute the right-hand side of (\ref{xindot}) into (\ref{xnxidot}) to obtain the {solvable} system of ODEs
\end{itemize}
\begin{subequations}
\begin{eqnarray}
&&\dot{x}_1=-\left[
(m_1+1)! \prod_{\ell=2}^N(x_1-x_\ell)
\right]^{-1}\notag\\
&&\cdot \sum_{j=1}^N (N-j+1)_{m_1} (x_1)^{N-j} f_j^{(1)}(y_1, \ldots, y_N),\label{x1dotSystem}\\
&&\dot{x}_n=-\left[\prod_{\ell=1, \ell \neq n}^N (x_n-x_\ell)^{-1} \right]
\sum_{m=1}^{N} g_m^{(1)}(y_1, \ldots, y_N, x_1) (x_n)^{N-m},  \label{xndotSystem}\\
&&2\leq n\leq N,\notag\\
&&y_j=(-1)^j \sigma_j^{(N+m_1)}(\underbrace{x_1, \ldots, x_1}_{m_1 \mbox{ times}}, x_1, \ldots, x_N),\; 1\leq j\leq N.
\end{eqnarray}
 \label{xndotSolvable}
 \end{subequations}
\begin{remark} Note that equation~\eqref{xndotSystem} with $n=1$ is equivalent to equation~\eqref{x1dotSystem} because  $\dot{x}_1$ is given by~\eqref{xnxidot} with $n=1$ as well as by~\eqref{x1dot}. The last observation implies that
$$
\sum_{j=1}^N (x_1)^{N-j} \dot{\xi}_j=\sum_{j=1}^N\frac{(N-j+1)_{m_1}}{(m_1+1)!}(x_1)^{N-j} \dot{y}_j.
$$
\end{remark}

By accomplishing the three steps listed above, we obtain system~\eqref{xndotSolvableExplicit} of Theorem~\eqref{Theorem1stOrder}.

The proof of solvability of system~\eqref{xndotSolvableExplicit} is instructive because it provides a process for solving the system.
Consider  system~\eqref{xndotSolvableExplicit} together with the initial conditions
\begin{equation}
x_n(t_0)=x_n^{(0)},
\end{equation}
where $t_0\in \mathbb{R}$, an initial value problem. The solution $(x_1(t), \ldots, x_N(t))$ of the last IVP at $t>t_0$ can be found as follows.
\begin{itemize}
\item[\textit{Step 1.}] Compute the corresponding initial conditions for system~\eqref{ydot_SystemThm}:
\begin{equation}
y_n^{(0)}=(-1)^n \sigma_n^{(N+m_1)}(\underbrace{x_1^{(0)}, \ldots, x_1^{(0)}}_{m_1 \mbox{ times}}, x_1^{(0)}, \ldots, x_N^{(0)}), \; 1\leq n \leq N.
\end{equation} 
\item[\textit{Step 2.}] Solve system~\eqref{ydot_SystemThm} with the initial conditions $$y_n(t_0)=y_n^{(0)}, 1\leq n \leq N$$
to obtain $y_1(t), \ldots, y_N(t)$.

\item[\textit{Step 3.}] Using the values of $y_1(t), \ldots, y_N(t)$ found on Step 2, solve polynomial equation~\eqref{PolyEqnx1}. Denote by $x_1(t)$ the solution of~\eqref{PolyEqnx1} that can be traced to the initial condition $x_1(t_0)=x_1^{(0)}$ by continuity.

\item[\textit{Step 4.}] Using the values of $y_1(t), \ldots, y_N(t), x_1(t)$ found on Steps~2 and~3, compute $y_{N+1}(t), \ldots, y_{N+m_1}(t)$ using formulas~\eqref{eq:y_Npk}.

\item[\textit{Step 5.}] Using the values of $y_1(t), \ldots, y_N(t), y_{N+1}(t), \ldots,y_{N+m_1}(t)$ found on Steps~2 and~4, find the roots of the polynomial
 $$p_{N+m_1}(z;t)=z^{N+m_1}+\sum_{n=1}^{N+m_1} y_n z^{N+m_1-n},$$
 see~\eqref{MainPolyy}. Note that $x_1(t)$ found on Step 3 is the root of multiplicity $(m_1+1)$ of the last polynomial. Assign the order of the remaining roots $x_2(t), \ldots, x_N(t)$ to ensure continuity of the functions $x_j(\tau)$ for $\tau \in[t_0,t]$, $2\leq j \leq N$.
\end{itemize}

A  {solvable second-order} nonlinear system of ODEs for $x_1, \ldots, x_N$ can be obtained in a similar manner, by accomplishing the following steps.
\begin{itemize}
\item[\textit{Step 1.}] Assign a {solvable} evolution of $\vec{y}^{(N)}=(y_1, \ldots, y_N)$ via a system
\begin{eqnarray}
&&\ddot{y}_n=f^{(2)}_n(y_1, \ldots, y_N, \dot{y}_1, \ldots, \dot{y}_N), \label{System_yddot}\\
&&1\leq n \leq N.\notag
\end{eqnarray}

\item[\textit{Step 2.}] Express $\ddot{\xi}_1, \ldots, \ddot{\xi}_N$ in terms of $y_1, \ldots, y_N$, $\dot{y}_1, \ldots, \dot{y}_N$, $x_1$, $\dot{x}_1$ and $\ddot{x}_1$ by using formulas (\ref{XiInTOy}) with $\ddot{y}_n=f^{(2)}_n(y_1, \ldots, y_N, \dot{y}_1, \ldots, \dot{y}_N)$ to obtain a formula
\begin{eqnarray}
&&\ddot{\xi}_n=\tilde{g}^{(2)}_n(y_1, \ldots, y_N, \dot{y}_1, \ldots, \dot{y}_N, x_1, \dot{x}_1,\ddot{x}_1), \label{xinddotPrelim}\\
&&1\leq n \leq N.\notag
\end{eqnarray}

\item[\textit{Step 3.}] Plug in $\ddot{x}_1$ given by~\eqref{x1ddot} with $\ddot{y}_n=f_n^{(2)}(\vec{y}^{(N)}, \dot{\vec{y}}^{(N)})$ into~\eqref{xinddotPrelim} to obtain  formulas
\begin{eqnarray}
&&\ddot{\xi}_n=g^{(2)}_n(\vec{x}, \dot{\vec{x}}, \vec{y}^{(N)}, \dot{\vec{y}}^{(N)}), \label{xinddot}\\
&&1\leq n \leq N.\notag
\end{eqnarray}

\item[\textit{Step 4.}] Substitute the right-hand side of (\ref{xinddot}) into (\ref{xnxiddot}) to obtain the {solvable} system of ODEs
\begin{subequations}
\begin{eqnarray}
&&\ddot{x}_1=-\left[ (m_1+1)! \prod_{n=2}^N (x_1-x_n) \right]^{-1} \sum_{j=1}^N (N-j+1)_{m_1} (x_1)^{N-j} f_j^{(2)}(\vec{y}^{(N)}, \dot{\vec{y}}^{(N)})\notag\\
&& +\dot{x}_1 \sum_{n=2}^N \frac{m_1\dot{x}_1 +2 \dot{x}_n}{x_1-x_n},
 \label{x1ddotSolvable}
\\
&&\ddot{x}_n=\sum_{\ell=1, \ell \neq n}^N \frac{2 \dot{x}_n \dot{x}_\ell}{x_n-x_\ell} \notag\\
&&-
\left[\prod_{\ell=1, \ell \neq n}^N (x_n-x_\ell)^{-1} \right]
\sum_{m=1}^N (x_n)^{N-m} 
g^{(2)}_m(\vec{x}, \dot{\vec{x}}, \vec{y}^{(N)}, \dot{\vec{y}}^{(N)})
, \;\; 2\leq n\leq N,
 \label{xnddotSolvable} \\
&&y_j=(-1)^j \sigma_j^{(N+m_1)}(\underbrace{x_1, \ldots, x_1}_{m_1 \mbox{ times}}, x_1, \ldots, x_N),\notag\\
&&
\dot{y}_j=(-1)^j \sum_{k=1}^N \left\{ \frac{\partial }{\partial x_k}\left[\sigma_j^{(N+m_1)}(\underbrace{x_1, \ldots, x_1}_{m_1 \mbox{ times}}, x_1, \ldots, x_N) \right] \dot{x}_k \right\}
\notag\\
&&1\leq j\leq N. \notag
\end{eqnarray}
\end{subequations}
\end{itemize}

\begin{remark} Note that equation~\eqref{xnddotSolvable} with $n=1$ is equivalent to equation~\eqref{x1ddotSolvable} because  $\ddot{x}_1$ is given by~\eqref{xnxiddot} with $n=1$ as well as by~\eqref{x1ddot}. The last observation implies that
$$
\sum_{j=1}^N (x_1)^{N-j} \ddot{\xi}_j=\sum_{j=1}^N\frac{(N-j+1)_{m_1}}{(m_1+1)!}(x_1)^{N-j} \ddot{y}_j
-m_1(\dot{x}_1)^2 \sum_{k=2}^N \left[ \prod_{\ell=2, \ell \neq k}^N (x_1-x_\ell) \right].
$$
\end{remark}

By accomplishing the four steps listed above, we obtain system~\eqref{xnddotSolvableExplicit} of Theorem~\ref{Theorem2ndOrder}. Let us prove that this system is solvable.

Consider  system~\eqref{xnddotSolvableExplicit} together with the initial conditions
\begin{equation}
x_n(t_0)=x_n^{(0)}, \;\;\; \dot{x}_n(t_0)=x_n^{(1)},
\end{equation}
where $t_0\in \mathbb{R}$, an initial value problem. The solution $(x_1(t), \ldots, x_N(t))$ of the last IVP at $t>t_0$ can be found as follows.
\begin{itemize}

\item[\textit{Step 1.}] Compute the corresponding initial conditions for system~\eqref{ydot_SystemThm}:
\begin{eqnarray}
&&y_n^{(0)}=(-1)^n \sigma_n^{(N+m_1)}(\underbrace{x_1^{(0)}, \ldots, x_1^{(0)}}_{m_1 \mbox{ times}}, x_1^{(0)}, \ldots, x_N^{(0)}), 
\notag\\
&&{y}_n^{(1)}=(-1)^n \sum_{j=1}^N \left\{ \frac{\partial }{\partial x_j}\left[\sigma_n^{(N+m_1)}(\underbrace{x_1, \ldots, x_1}_{m_1 \mbox{ times}}, x_1, \ldots, x_N) \right]_{\Big|_{x_j=x_j^{(0)}}} x_j^{(1)} \right\} \notag\\
&& 1\leq n \leq N.
\end{eqnarray} 

\item[\textit{Step 2.}] Solve system~\eqref{yddot_SystemThm} with the initial conditions $$y_n(t_0)=y_n^{(0)},\; 
\dot{y}_n(t_0)=y_n^{(1)},\;\; 1\leq n \leq N,$$
to obtain $y_1(t), \ldots, y_N(t)$.

\item[\textit{Step 3.}] Using the values of $y_1(t), \ldots, y_N(t)$ found on Step 2, solve polynomial equation~\eqref{PolyEqnx1}. Denote by $x_1(t)$ the solution of~\eqref{PolyEqnx1} that can be traced to the initial conditions $x_1(t_0)=x_1^{(0)}, 
\dot{x}_1(t_0)=x_1^{(1)}$ by continuity.

\item[\textit{Step 4.}] Find $x_2(t), \ldots, x_N(t)$ by executing Steps~4 and~5 from the proof of Theorem~\ref{Theorem1stOrder}.

\end{itemize}

\section{Examples of  solvable $N$-body problems}
\label{Sect3}

In this section we apply Theorem~\ref{Theorem2ndOrder} to construct several solvable $2$- or $3$-body problems. To the best of our knowledge, these examples are new.

\subsection{Two-body problems}
 Consider the solvable $N$-body problem~\eqref{xnddotSolvableExplicit} of Theorem~\ref{Theorem2ndOrder} with $N=2$. In this case, polynomial~\eqref{MainPoly} reduces to
\begin{eqnarray}
p_{2+m_1}(z;t)=(z-x_1)^{m_1+1}(z-x_2)=z^{m_1+2}+\sum_{n=1}^{m_1+2} y_n z^{m_1+2-n}
\label{MailPolyEx1}
\end{eqnarray}
so that
\begin{eqnarray}
&&y_1=-(m_1+1) x_1-x_2, \; y_2=\frac{(m_1+1) }{2} x_1 (m_1 x_1+2 x_2 );\label{y12Ex1}\\
&&\dot{y}_1=-(m_1+1) \dot{x}_1-\dot{x}_2, \;
\dot{y}_2=(m_1+1)(\dot{x}_1 x_2+ x_1 \dot{x}_2 +m_1 x_1 \dot{x}_1), \label{doty12Ex1}
\end{eqnarray}
see~\eqref{yjThm2},~\eqref{dotyjThm2}. As for the parameters $\vec{\gamma}$ and $\underline{\alpha}$ of system~\eqref{xnddotSolvableExplicit}, we will need the following coefficients:
\begin{eqnarray*}
&&\alpha_{21}=m_1,\\
&&\gamma_1=-m_1, \; \gamma_2=- \binom{m_1+1}{2},\\
\end{eqnarray*}
see~\eqref{alphabetanm},~\eqref{gamman}.
Thus, in the case where $N=2$ system~\eqref{xnddotSolvableExplicit} reduces to
\begin{eqnarray}
&&\ddot{x}_1=-\frac{1}{(m_1+1)(x_1-x_2)}\left[ (m_1+1) x_1 f_1^{(2)}(\vec{y}^{(2)}, \dot{\vec{y}}^{(2)})+f_2^{(2)}(\vec{y}^{(2)}, \dot{\vec{y}}^{(2)})\right]\notag\\
&& +\dot{x}_1\frac{m_1\dot{x}_1+2\dot{x}_2}{x_1-x_2},\notag\\
&&\ddot{x}_2=\frac{1}{x_1-x_2}\Big[(m_1 x_1+x_2) f_1^{(2)}(\vec{y}^{(2)}, \dot{\vec{y}}^{(2)})+f_2^{(2)} (\vec{y}^{(2)}, \dot{\vec{y}}^{(2)})\Big]\notag\\
&&- (m_1+1)\dot{x}_1 \frac{m_1\dot{x}_1+2 \dot{x}_2 }{x_1-x_2},
\label{2BodyGeneral}
\end{eqnarray}
where $\vec{y}^{(2)}=(y_1, y_2)$ and $\dot{\vec{y}}^{(2)}=(\dot{y}_1, \dot{y}_2)$ are given by~\eqref{y12Ex1} and~\eqref{doty12Ex1}. By Theorem~\ref{Theorem2ndOrder}, the last $2$-body problem is algebraically solvable if system~\eqref{yddot_SystemThm} with $N=2$ is algebraically solvable.

\begin{remark} \label{rem:m11_2}
Note that in the special case where $m_1=1$, that is, if $x_1$ is the root of polynomial~\eqref{MailPolyEx1} of multiplicity~2, the last $2$-body problem coincides with system~(15) reported in~\cite{23}. 
\end{remark}

\noindent \textbf{Example 3.1.1.} In this example, we consider the following  generating model as a point of departure:
\begin{eqnarray}
\ddot{y}_1=\mathbf{i}\, r_1\, \omega\, \dot{y}_1,\notag\\
\ddot{y}_2=\mathbf{i} \,r_2 \,\omega\, \dot{y}_2,
\label{Model3.1.1}
\end{eqnarray}
where $\omega$ is a nonvanishing real number, $r_1, r_2$ are nonvanishing rational numbers and $\mathbf{i}$ is the imaginary unit, so that $\mathbf{i}^2=-1$.
This system is Hamiltonian and integrable; its solution can be found explicitly:
\begin{eqnarray}
y_m(t)=y_m(0)+\dot{y}_m(0)\Big[ \frac{\exp(\mathbf{i}\, r_m \,\omega\, t)-1}{\mathbf{i}\, r_m \,\omega} \Big],~m=1,2.
\label{ySolnModel3.1.1}
\end{eqnarray}
It is clear that the last solution is isochronous with a period $T$ that is an integer multiple of $2\pi/|\omega|$.

Via Theorem~\ref{Theorem2ndOrder}, model~\eqref{Model3.1.1} generates the following solvable $2$-body problem, see~\eqref{2BodyGeneral}:
\begin{eqnarray}
&&\ddot{x}_1=\frac{\dot{x}_1(m_1 \dot{x}_1+ 2\dot{x}_2)}{x_1-x_2}\notag\\
&&+\mathbf{i}\, \omega \left[\frac{-r_2 \dot{x}_1 x_2+(r_1+m_1 r_1-m_1 r_2) x_1 \dot{x}_1+(r_1-r_2)x_1 \dot{x}_2}{x_1-x_2}\right]
\notag\\
&&\ddot{x}_2=-(m_1+1) \dot{x}_1 \left[ \frac{m_1 \dot{x}_1+ 2 \dot{x}_2}{x_1-x_2} \right]\notag\\
&&- \frac{\mathbf{i}\, \omega}{x_1-x_2} \Big[
 m_1 (m_1+1) (r_1-r_2) x_1 \dot{x}_1+(m_1(r_1-r_2)-r_2) x_1 \dot{x}_2\notag\\
&&+(1+m_1)(r_1-r_2) \dot{x}_1 x_2 + r_1 x_2 \dot{x}_2
\Big].
\label{2Body3.1.1}
\end{eqnarray}
The last $2$-body problem is isochronous. Indeed, each solution $y_1(t), y_2(t)$ of the generating model~\eqref{Model3.1.1} is periodic with the same period $T$. Because the remaining coefficients $y_3(t), \ldots, y_{2+m_1}(t)$  of the polynomial $p_{N+m_1}(z;t)$ with $N=2$, see~\eqref{MainPoly}, are expressed in terms of $y_1(t), y_2(t)$ via~\eqref{eq:y_Npk} with $N=2$, they are also periodic with the same period $T$. But then the zeros $x_1(t), x_2(t)$ of the polynomial  $p_{N+m_1}(z;t)$ with $N=2$ are periodic the period $T$, or possibly an integer multiple of $T$ (due to the possibility of the zeros exchanging their role at collisions, see~\cite{25}).

In Figures~\ref{Ex311F1},~\ref{Ex311F2},~\ref{Ex311F3},~\ref{Ex311F4} we provide the plots of the solutions of system~(\ref{2Body3.1.1}) with the parameters
\begin{equation}
m_1=17;\;\; r_1=\frac{1}{2},\; \; r_2=\frac{1}{3}, \;\;\omega=2\pi, 
\label{par:2Body3.1.1}
\end{equation}
satisfying the initial conditions
\begin{eqnarray}
&&x_1(0) =3.19 + 3.67 \; \mathbf{i}, \;\;\;\;\;\;\;\;\,x_1'(0) =0.56 + 4.97 \; \mathbf{i}, \notag\\
&&x_2(0) =-47.46 - 23.83 \; \mathbf{i}, \;\;x_2'(0) =27.85 - 52.55 \; \mathbf{i}.
\label{InitCond:2Body3.1.1}
\end{eqnarray}

\begin{minipage}{\linewidth}
      \centering
      \begin{minipage}{0.45\linewidth}
          \begin{figure}[H]
              \includegraphics[width=\linewidth]{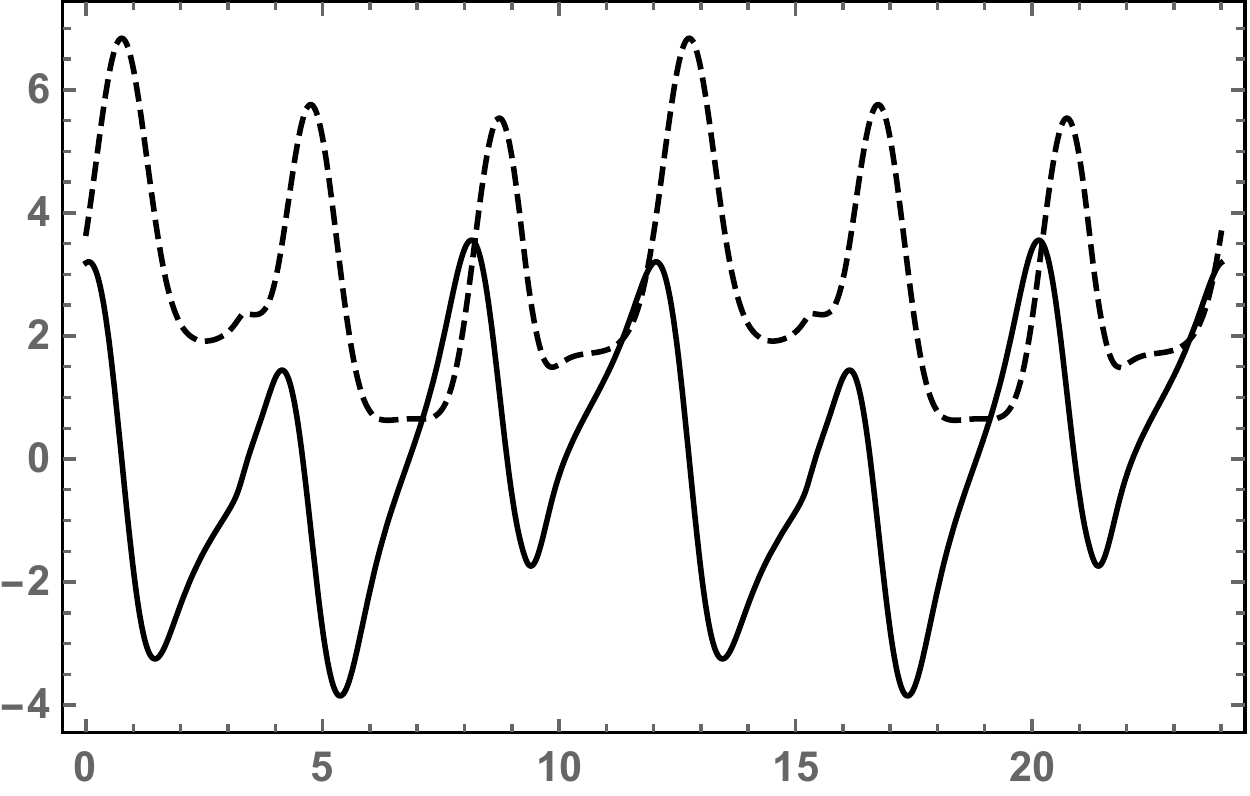}
              \caption{Initial value problem~(\ref{2Body3.1.1}),~(\ref{par:2Body3.1.1}),~(\ref{InitCond:2Body3.1.1}). Graphs of the real (bold curve) and imaginary 
              (dashed curve) parts of the coordinate $x_1(t)$; period $12$.}
              \label{Ex311F1}
          \end{figure}
      \end{minipage}
      \hspace{0.05\linewidth}
      \begin{minipage}{0.45\linewidth}
          \begin{figure}[H]
              \includegraphics[width=\linewidth]{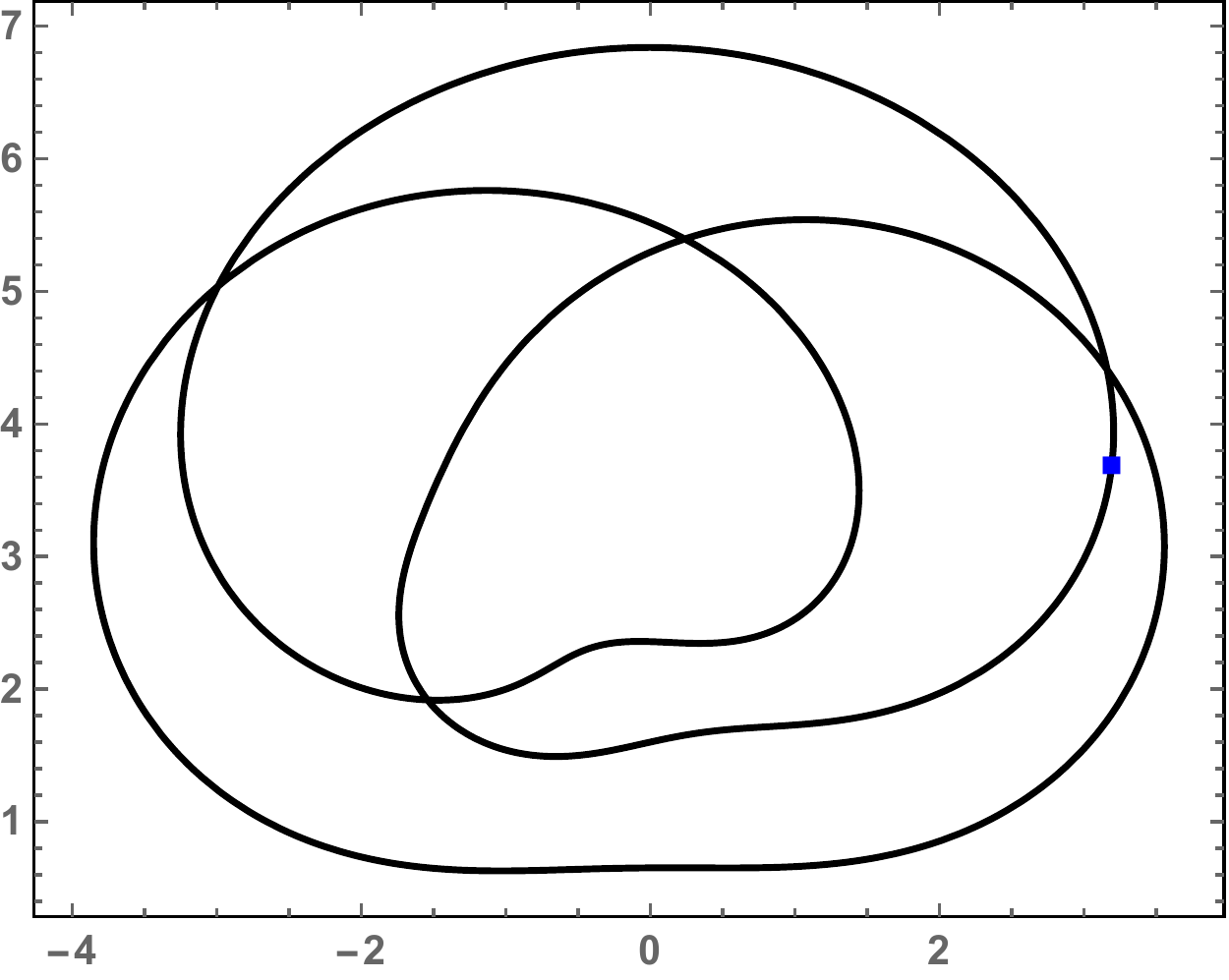}
              \caption{Initial value problem~(\ref{2Body3.1.1}),~(\ref{par:2Body3.1.1}),~(\ref{InitCond:2Body3.1.1}). Trajectory, in the complex $x$-plane, of  $x_1(t)$; 
              period $12$. The   square indicates the initial condition $x_1(0)=3.19 + 3.67 \; \mathbf{i}$.}
              \label{Ex311F2}
          \end{figure}
      \end{minipage}
  \end{minipage}
  
  \begin{minipage}{\linewidth}
      \centering
      \begin{minipage}{0.45\linewidth}
          \begin{figure}[H]
              \includegraphics[width=\linewidth]{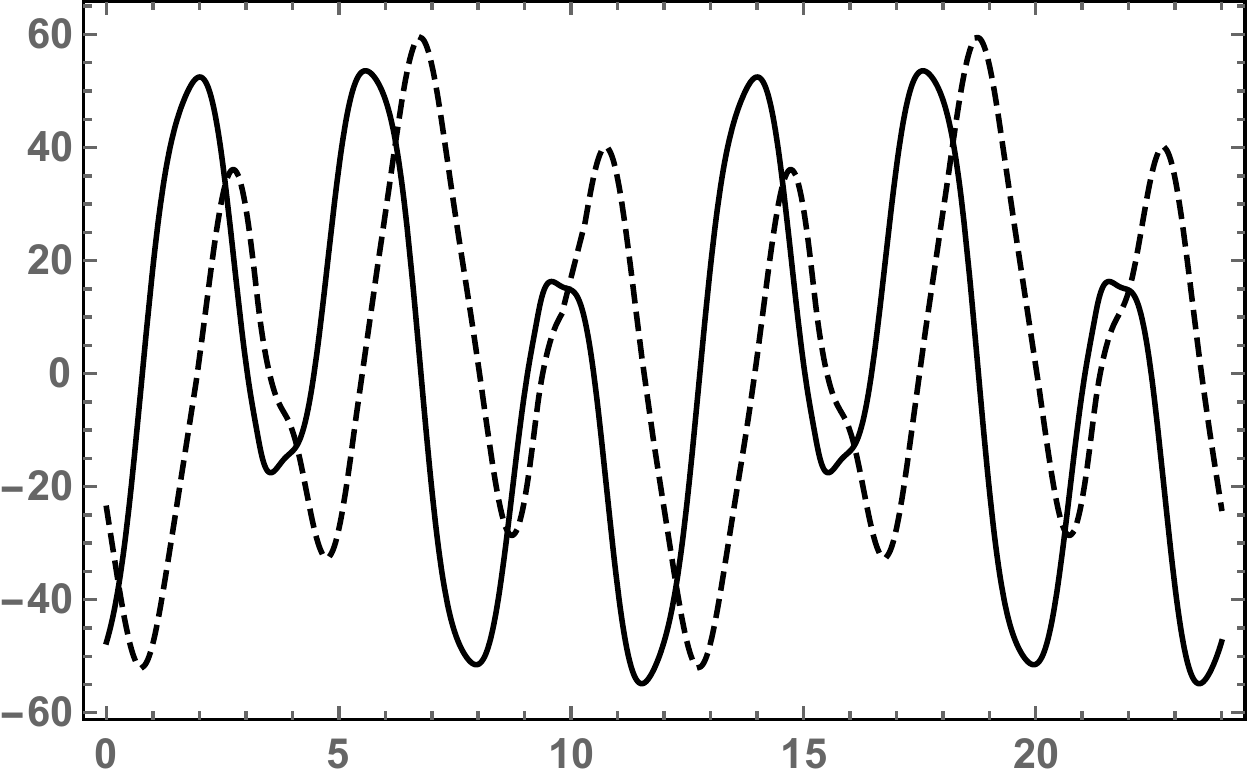}
              \caption{Initial value problem~(\ref{2Body3.1.1}),~(\ref{par:2Body3.1.1}),~(\ref{InitCond:2Body3.1.1}).  Graphs of the real (bold curve) and imaginary 
              (dashed curve) parts of the coordinate $x_2(t)$; period $12$.}
              \label{Ex311F3}
          \end{figure}
      \end{minipage}
      \hspace{0.05\linewidth}
      \begin{minipage}{0.45\linewidth}
          \begin{figure}[H]
              \includegraphics[width=\linewidth]{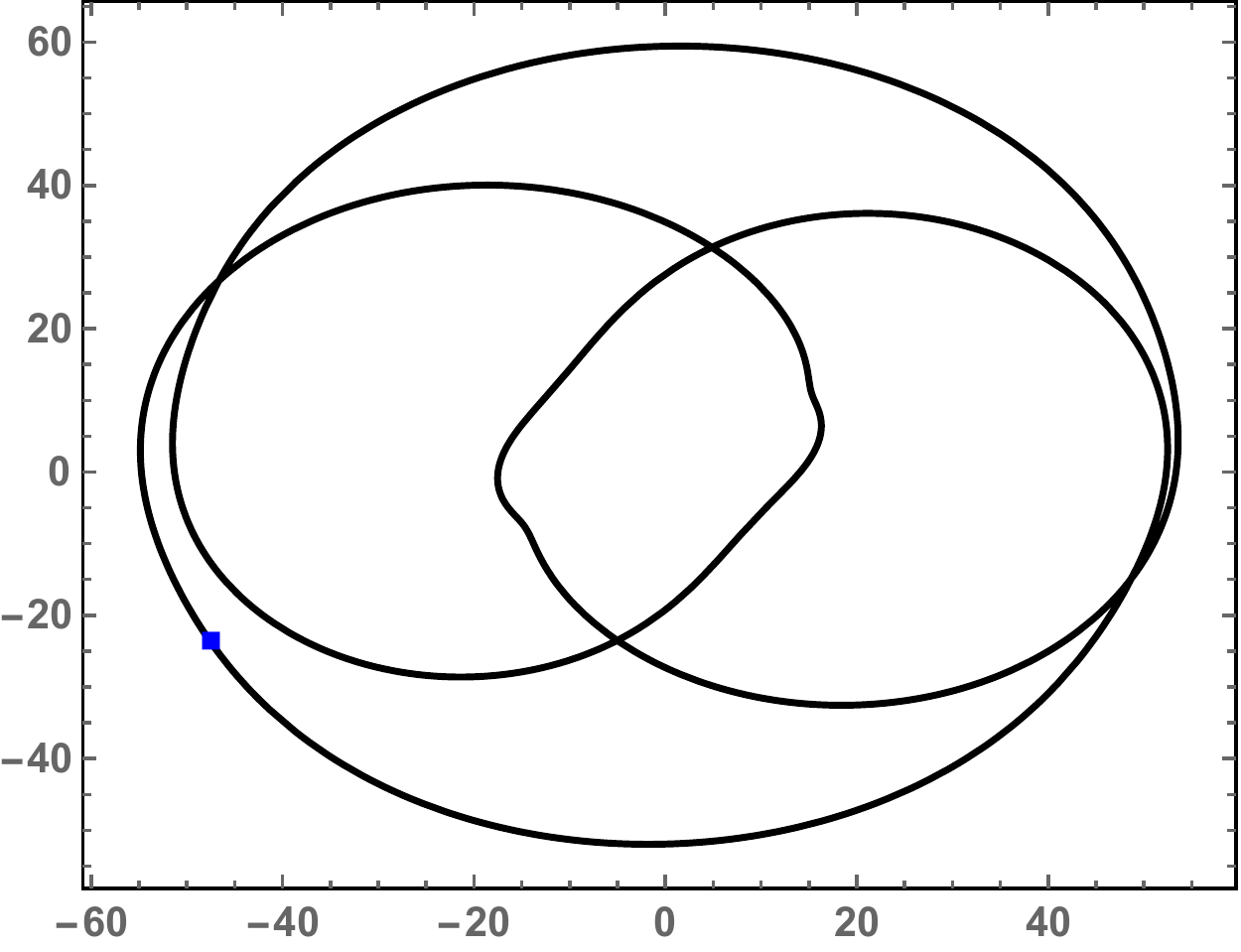}
              \caption{Initial value problem~(\ref{2Body3.1.1}),~(\ref{par:2Body3.1.1}),~(\ref{InitCond:2Body3.1.1}). Trajectory, in the complex $x$-plane, of  $x_2(t)$; 
              period $12$. The   square indicates the initial condition $x_2(0)= -47.46 - 23.83 \; \mathbf{i}$.}
              \label{Ex311F4}
          \end{figure}
      \end{minipage}
  \end{minipage}
\smallskip

\begin{remark}
\label{Remark 3.1.1} The period of the solution of the generating model~\eqref{Model3.1.1} is $6$, however, the period of the solution of the initial value problem~(\ref{2Body3.1.1}),~(\ref{par:2Body3.1.1}),~(\ref{InitCond:2Body3.1.1}) is $12$. This is due to the fact that the zeros of a polynomial with periodic coefficients may interchange their roles when they collide, see~\cite{25}.
\end{remark}

\begin{remark} \label{Remark 3.1.2} If $r_{1}=r_{2}=r,$ the $2$-body problem~(\ref{2Body3.1.1}) simplifies to the following system:
\begin{eqnarray}
&&\ddot{x}_1=\frac{\dot{x}_1(m_1 \dot{x}_1+ 2\dot{x}_2)}{x_1-x_2}+\mathbf{i}\, r\,\omega\, \dot{x}_1 
\notag\\
&&\ddot{x}_2=-(m_1+1)  \left[ \frac{\dot{x}_1(m_1 \dot{x}_1+ 2 \dot{x}_2)}{x_1-x_2} \right]+\mathbf{i}\,r\, \omega \,\dot{x}_2.\label{2Body3.1.1r1eqr2}
\end{eqnarray}
Note that in the case where $m_1=0$ and $r=1$, the last system reduces to the $2$-body goldfish model~\cite{7}.
\end{remark}

\noindent \textbf{Example 3.1.2.} In this example, the  generating model is
\begin{eqnarray}
\ddot{y}_1=- r_1^2\, \omega^2\, {y}_1,\notag\\
\ddot{y}_2=-r_2^2 \,\omega^2\, {y}_2,
\label{Model3.1.2}
\end{eqnarray}
where, as before, $\omega$ is a nonvanishing real number and $r_1, r_2$ are nonvanishing rational numbers.
System~\eqref{Model3.1.2} is Hamiltonian and integrable; its solution 
\begin{eqnarray}
y_m(t)=y_m(0)\cos(r_m \omega\, t)+\frac{1}{r_m \omega}\dot{y}_m(0) \sin(r_m \omega\, t),~m=1,2,
\label{ySolnModel3.1.2}
\end{eqnarray}
 is isochronous with a period that is an integer multiple of $2\pi/|\omega|$.

Via Theorem~\ref{Theorem2ndOrder}, model~\eqref{Model3.1.2} generates the following solvable $2$-body problem, see~\eqref{2BodyGeneral}:
\begin{eqnarray}
&&\ddot{x}_1= \frac{\dot{x}_1(m_1 \dot{x}_1+ 2 \dot{x}_2)}{x_1-x_2} \notag\\
&&+\frac{\left[ -2(m_1+1) r_1^2+m_1 r_2^2\right] \omega^2 x_1^2+2(r_2^2-r_1^2)\omega^2 x_1 x_2}{2(x_1-x_2)},
\notag\\
&&\ddot{x}_2=(m_1+1) \frac{\dot{x}_1(m_1 \dot{x}_1+ 2 \dot{x}_2)}{x_1-x_2} \notag\\
&&+\frac{1}{2(x_1-x_2)}\Big \{
m_1(m_1+1) (2 r_1^2-r_2^2) \omega^2 x_1^2\notag\\
&&
+2 \left[ (1+2 m_1) r_1^2 -(1+m_1) r_2^2\right] \omega^2 x_1 x_2-2 r_1^2 \omega^2 x_2^2
\Big\}.
\label{2Body3.1.2}
\end{eqnarray}
System~\eqref{2Body3.1.2} is isochronous for the same reasons that system~\eqref{2Body3.1.1} is isochronous, see the paragraph following display~\eqref{2Body3.1.1}.

In Figures~\ref{Ex312F1},~\ref{Ex312F2},~\ref{Ex312F3},~\ref{Ex312F4} we provide the plots of the solutions of system~(\ref{2Body3.1.2}) with the parameters
\begin{equation}
m_1=11;\;\; r_1=\frac{1}{3},\; \; r_2=\frac{1}{2}, \;\;\omega=2\pi, 
\label{par:2Body3.1.2}
\end{equation}
satisfying the initial conditions
\begin{eqnarray}
&&x_1(0) =-18.14 + 35.16 \;\mathbf{i}, \;\;x_1'(0) =51.09 - 77.17 \;\mathbf{i}, \notag\\
&&x_2(0) =102.58 - 154.58 \;\mathbf{i}, \;\;x_2'(0) =-308.45 + 508.99 \;\mathbf{i}.
\label{InitCond:2Body3.1.2}
\end{eqnarray}

\begin{minipage}{\linewidth}
      \centering
      \begin{minipage}{0.45\linewidth}
          \begin{figure}[H]
              \includegraphics[width=\linewidth]{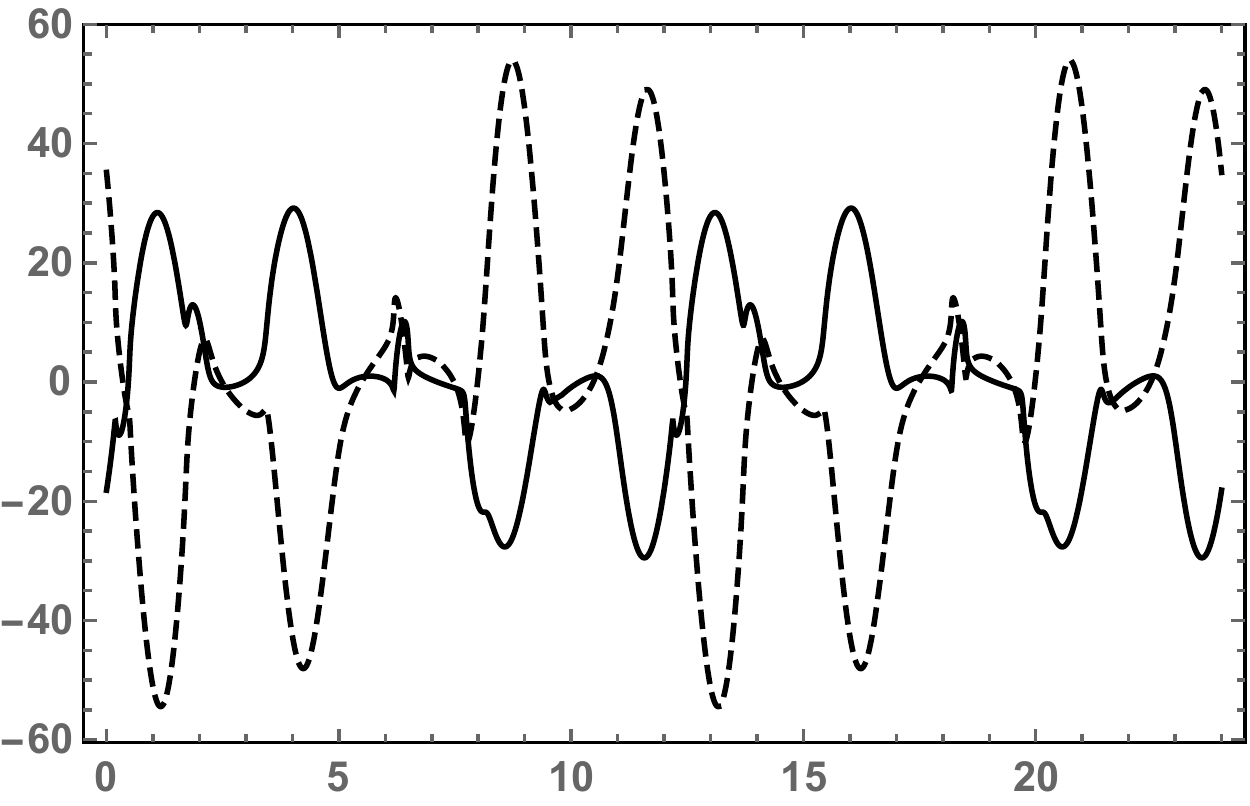}
              \caption{Initial value problem~(\ref{2Body3.1.2}),~(\ref{par:2Body3.1.2}),~(\ref{InitCond:2Body3.1.2}). Graphs of the real (bold curve) and imaginary 
              (dashed curve) parts of the coordinate $x_1(t)$; period $12$.}
              \label{Ex312F1}
          \end{figure}
      \end{minipage}
      \hspace{0.05\linewidth}
      \begin{minipage}{0.45\linewidth}
          \begin{figure}[H]
              \includegraphics[width=\linewidth]{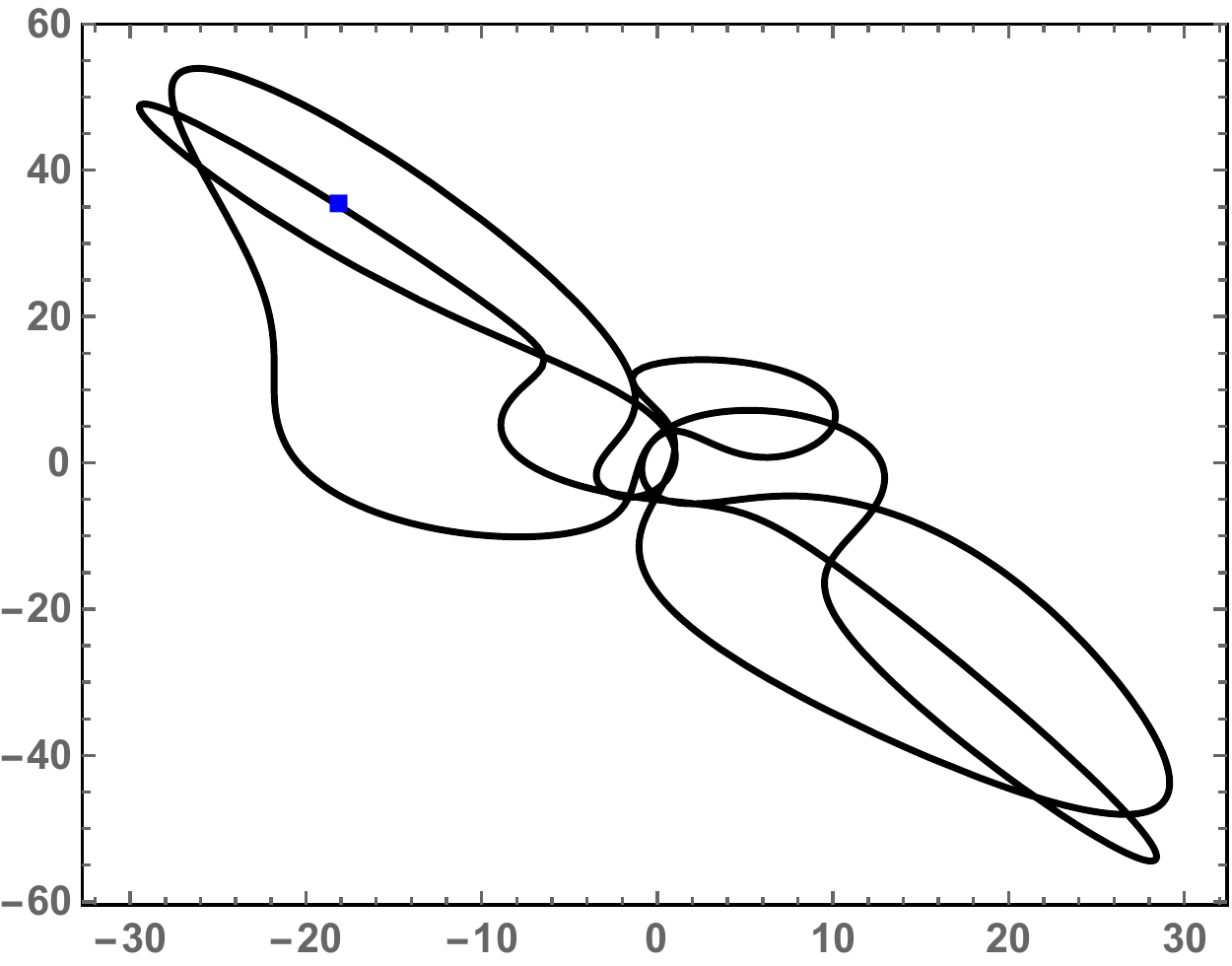}
              \caption{Initial value problem~(\ref{2Body3.1.2}),~(\ref{par:2Body3.1.2}),~(\ref{InitCond:2Body3.1.2}). Trajectory, in the complex $x$-plane, of  $x_1(t)$; 
              period $12$. The   square indicates the initial condition $x_1(0)=-18.14 + 35.16 \;\mathbf{i}$.}
              \label{Ex312F2}
          \end{figure}
      \end{minipage}
  \end{minipage}
  
  \begin{minipage}{\linewidth}
      \centering
      \begin{minipage}{0.45\linewidth}
          \begin{figure}[H]
              \includegraphics[width=\linewidth]{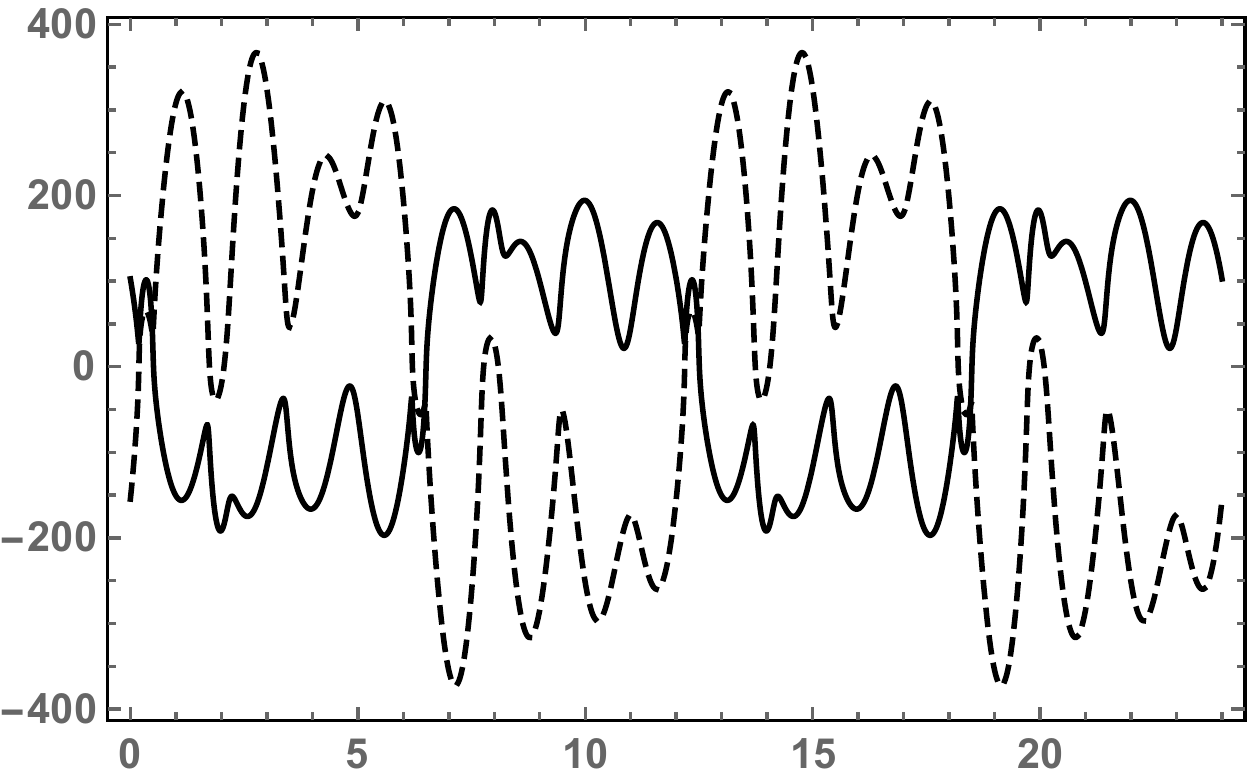}
              \caption{Initial value problem~(\ref{2Body3.1.2}),~(\ref{par:2Body3.1.2}),~(\ref{InitCond:2Body3.1.2}). Graphs of the real (bold curve) and imaginary 
              (dashed curve) parts of the coordinate $x_2(t)$; period $12$.}
              \label{Ex312F3}
          \end{figure}
      \end{minipage}
      \hspace{0.05\linewidth}
      \begin{minipage}{0.45\linewidth}
          \begin{figure}[H]
              \includegraphics[width=\linewidth]{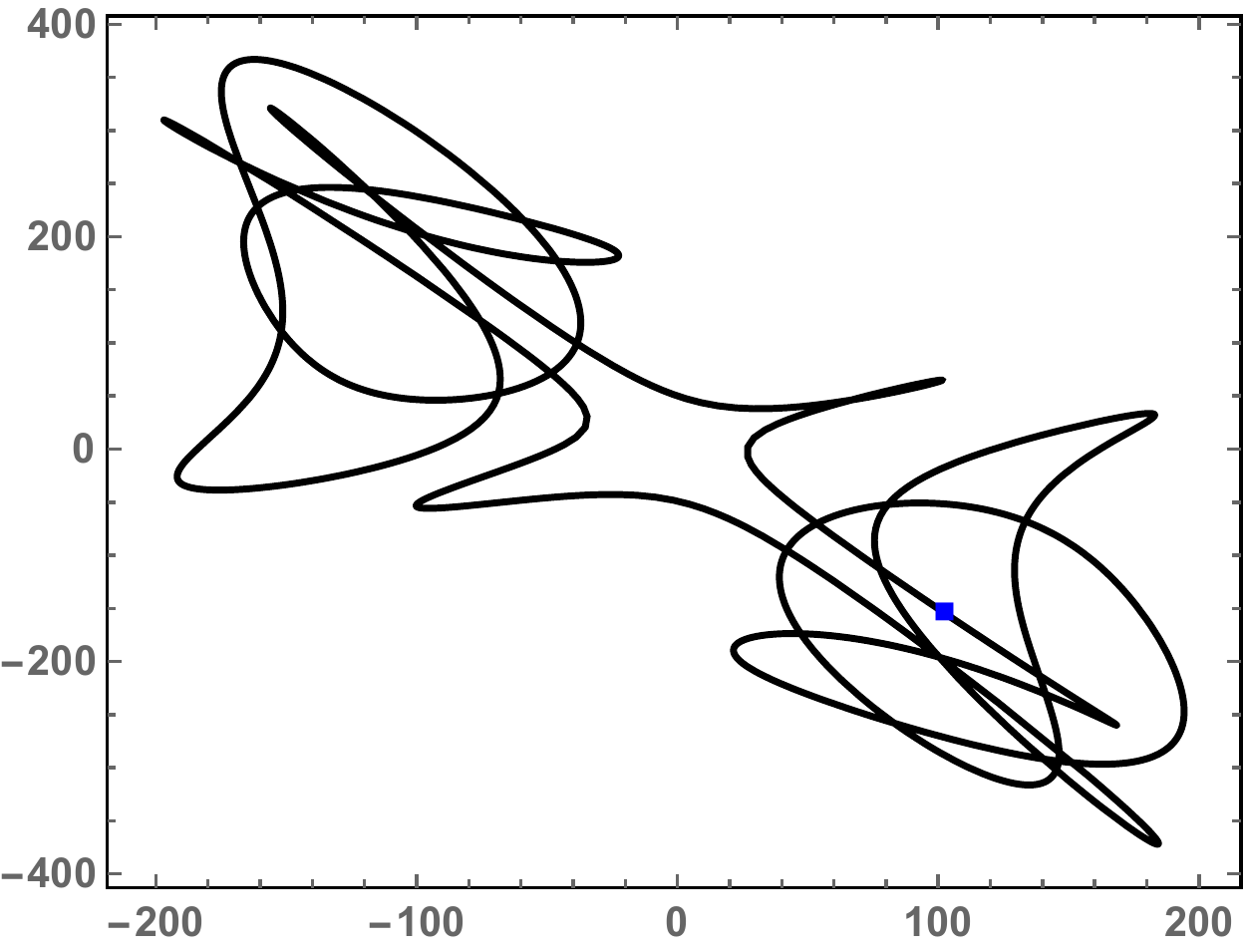}
              \caption{Initial value problem~(\ref{2Body3.1.2}),~(\ref{par:2Body3.1.2}),~(\ref{InitCond:2Body3.1.2}). Trajectory, in the complex $x$-plane, of  $x_2(t)$; 
              period $12$. The   square indicates the initial condition $x_2(0)=102.58 - 154.58 \;\mathbf{i}$.}
              \label{Ex312F4}
          \end{figure}
      \end{minipage}
  \end{minipage}
\smallskip

\noindent \textbf{Example 3.1.3.} In this example, the  generating model is
\begin{eqnarray}
&&\ddot{y}_1=\mathbf{i}\, r_1\, \omega\, \dot{y}_1,\notag\\
&&\ddot{y}_2=-r_2^2 \,\omega^2\, {y}_2,
\label{Model3.1.3}
\end{eqnarray}
where, as before, $\omega$ is a nonvanishing real number and $r_1, r_2$ are nonvanishing rational numbers.
System~\eqref{Model3.1.3} is Hamiltonian and integrable; its solution 
\begin{eqnarray}
&&y_1(t)=y_1(0)+\dot{y}_1(0)\Big[ \frac{\exp(\mathbf{i}\, r_1 \,\omega\, t)-1}{\mathbf{i}\, r_1 \,\omega} \Big],\notag\\
&&y_2(t)=y_2(0)\cos(r_2\, \omega\, t)+\frac{1}{r_2 \omega}\dot{y}_2(0) \sin(r_2\, \omega\, t),
\end{eqnarray}
 is isochronous with a period that is an integer multiple of $2\pi/|\omega|$.

Via Theorem~\ref{Theorem2ndOrder}, model~\eqref{Model3.1.3} generates the following solvable $2$-body problem, see~\eqref{2BodyGeneral}:
\begin{eqnarray}
&&\ddot{x}_1= \frac{\dot{x}_1(m_1 \dot{x}_1+ 2 \,\dot{x}_2)}{x_1-x_2}+ 
\mathbf{i}\, r_1\,  \omega\,\frac{x_1 \left[ (1+m_1)\, \dot{x}_1+\dot{x}_2 \right]}{x_1-x_2}\notag\\
&&+r_2^2\, \omega^2\,  \frac{x_1(m_1 x_1+2 \,x_2)}{2\,(x_1-x_2)},
\notag\\
&&\ddot{x}_2=-(m_1+1) \frac{\dot{x}_1(m_1 \dot{x}_1+2\, \dot{x}_2)}{x_1-x_2} 
- \mathbf{i}\, r_1 \, \omega\,\frac{(m_1 x_1+x_2) \left[ (m_1+1) \dot{x}_1+\dot{x}_2\right]}{x_1-x_2}\notag\\
&&-(m_1+1)\, r_2^2\, \omega^2\, \frac{x_1(m_1x_1+2 \,x_2)}{2\,(x_1-x_2)}.
\label{2Body3.1.3}
\end{eqnarray}
System~\eqref{2Body3.1.3} is isochronous for the same reasons that system~\eqref{2Body3.1.1} is isochronous, see the paragraph following display~\eqref{2Body3.1.1}.

In Figures~\ref{Ex313F1},~\ref{Ex313F2},~\ref{Ex313F3},~\ref{Ex313F4} we provide the plots of the solutions of system~(\ref{2Body3.1.3}) with the parameters
\begin{equation}
m_1=3;\;\; r_1=\frac{1}{3},\; \; r_2=\frac{1}{4}, \;\;\omega=2\pi, 
\label{par:2Body3.1.3}
\end{equation}
satisfying the initial conditions
\begin{eqnarray}
&&x_1(0) =33.68 + 30.30 \;\mathbf{i}, \;\;\;\;\;\;\;x_1'(0) =66.18 + 77.73 \;\mathbf{i}, \notag\\
&&x_2(0) =-160.42 - 84.73 \;\mathbf{i}, \;\;x_2'(0) =-474.40 - 227.29 \;\mathbf{i}.
\label{InitCond:2Body3.1.3}
\end{eqnarray}

\begin{minipage}{\linewidth}
      \centering
      \begin{minipage}{0.45\linewidth}
          \begin{figure}[H]
              \includegraphics[width=\linewidth]{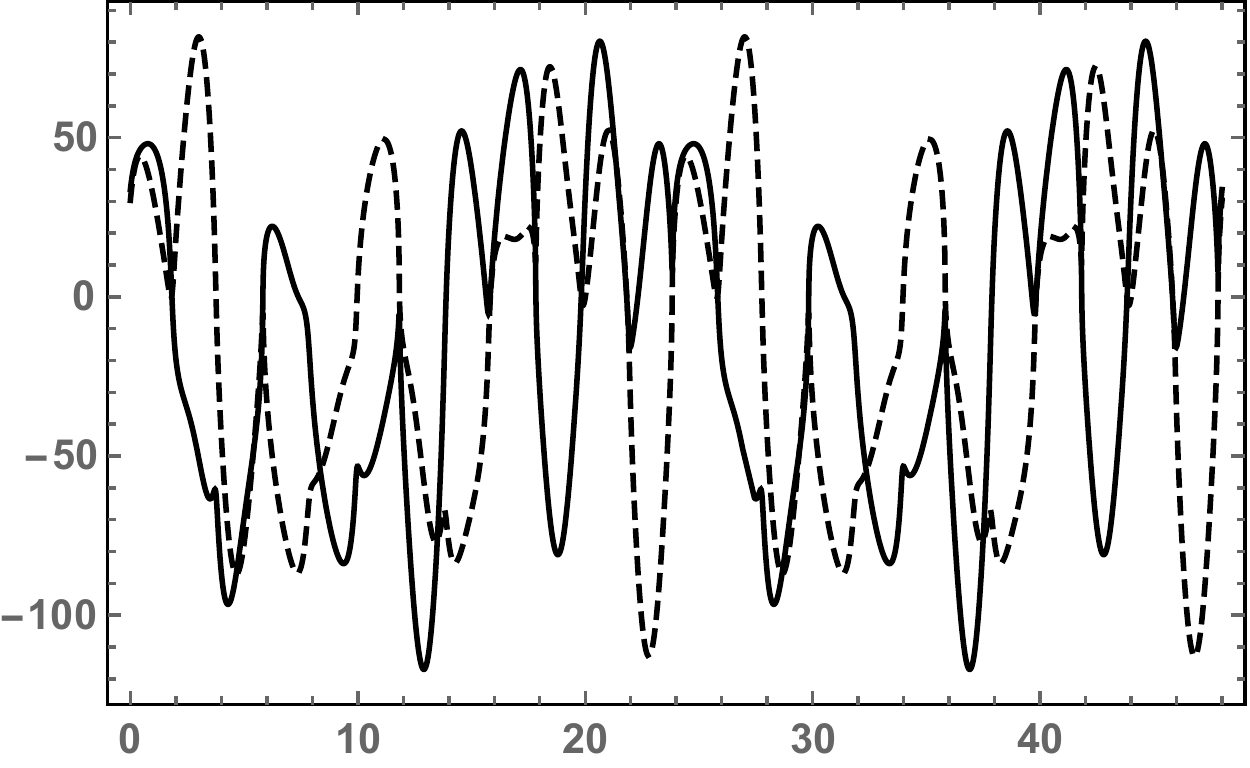}
              \caption{Initial value problem~(\ref{2Body3.1.3}),~(\ref{par:2Body3.1.3}),~(\ref{InitCond:2Body3.1.3}). Graphs of the real (bold curve) and imaginary 
              (dashed curve) parts of the coordinate $x_1(t)$; period $24$.}
              \label{Ex313F1}
          \end{figure}
      \end{minipage}
      \hspace{0.05\linewidth}
      \begin{minipage}{0.45\linewidth}
          \begin{figure}[H]
              \includegraphics[width=\linewidth]{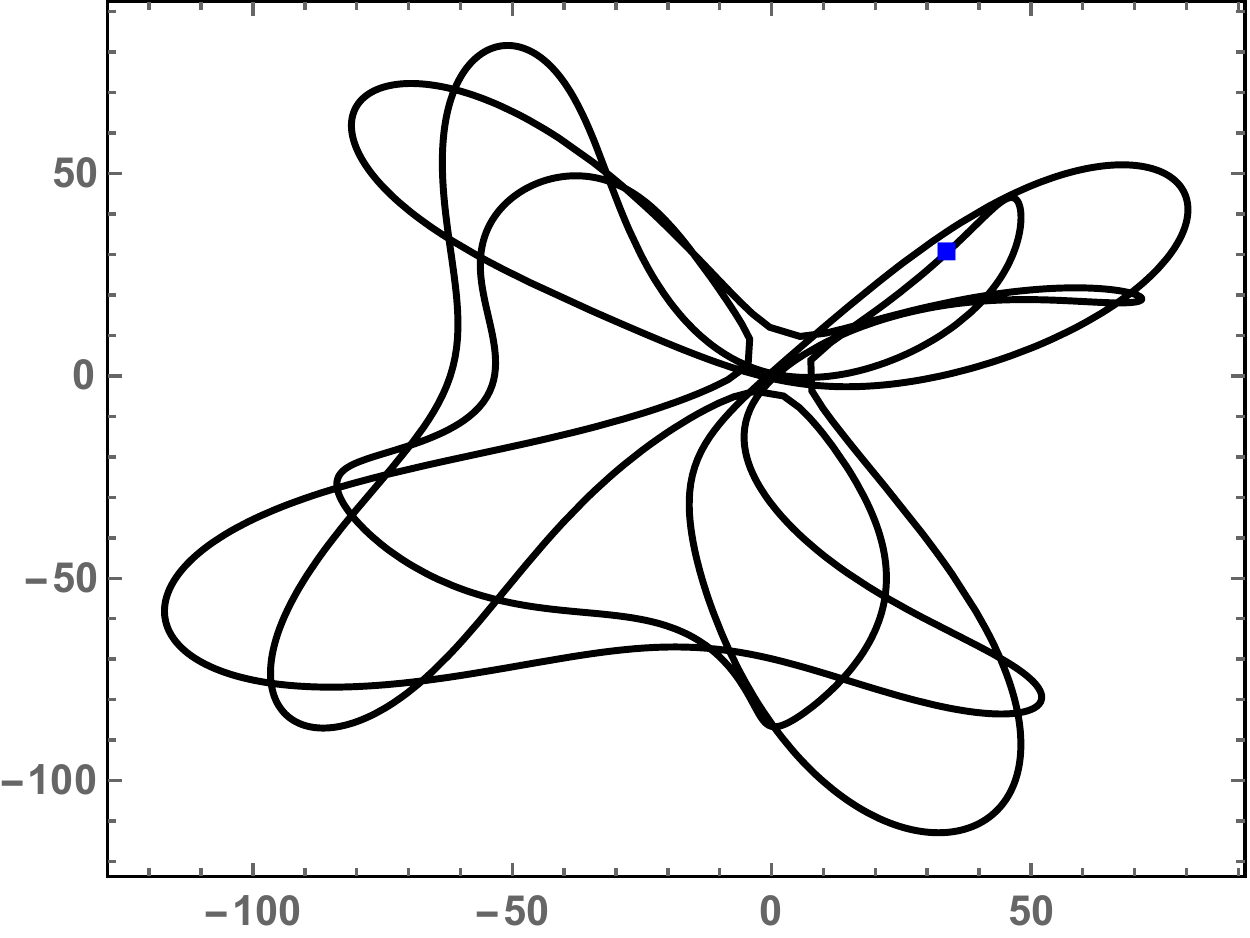}
              \caption{Initial value problem~(\ref{2Body3.1.3}),~(\ref{par:2Body3.1.3}),~(\ref{InitCond:2Body3.1.3}). Trajectory, in the complex $x$-plane, of  $x_1(t)$; 
              period $24$. The   square indicates the initial condition $x_1(0)=33.68 + 30.30 \;\mathbf{i}$.}
              \label{Ex313F2}
          \end{figure}
      \end{minipage}
  \end{minipage}
  
  \begin{minipage}{\linewidth}
      \centering
      \begin{minipage}{0.45\linewidth}
          \begin{figure}[H]
              \includegraphics[width=\linewidth]{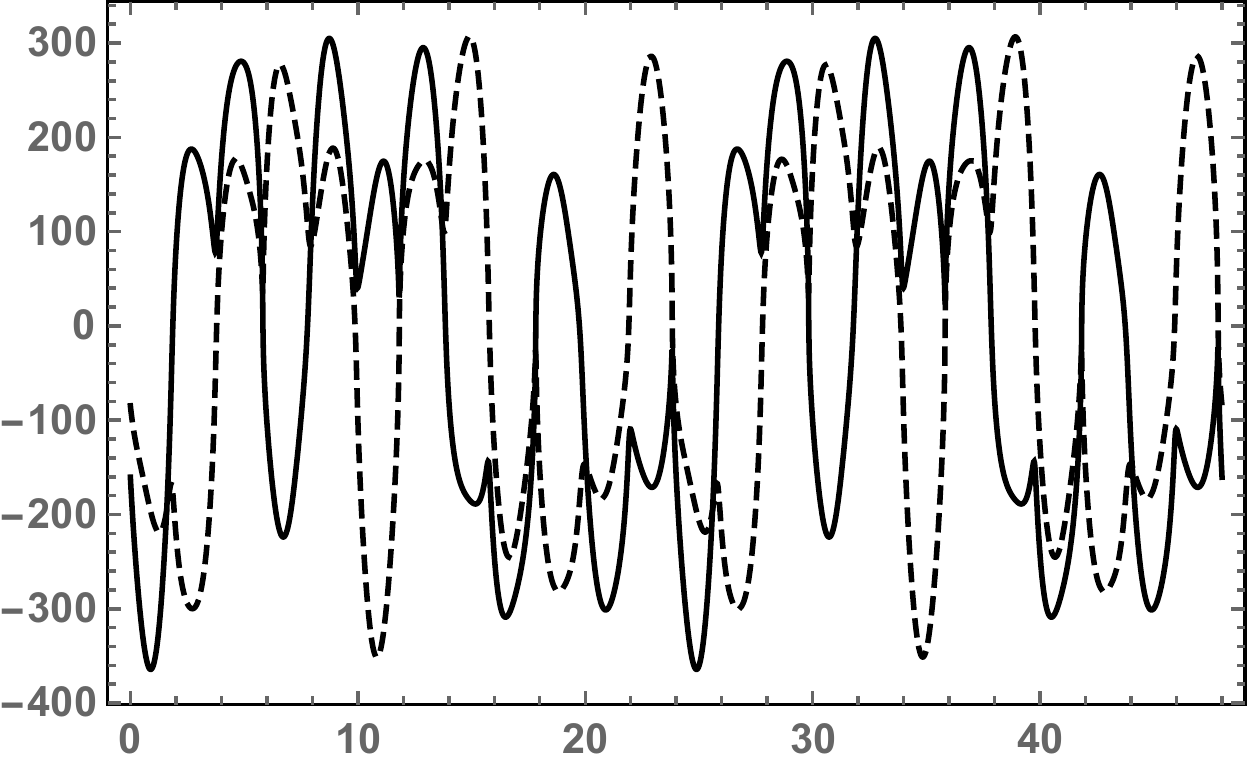}
              \caption{Initial value problem~(\ref{2Body3.1.3}),~(\ref{par:2Body3.1.3}),~(\ref{InitCond:2Body3.1.3}). Graphs of the real (bold curve) and imaginary 
              (dashed curve) parts of the coordinate $x_2(t)$; period $24$.}
              \label{Ex313F3}
          \end{figure}
      \end{minipage}
      \hspace{0.05\linewidth}
      \begin{minipage}{0.45\linewidth}
          \begin{figure}[H]
              \includegraphics[width=\linewidth]{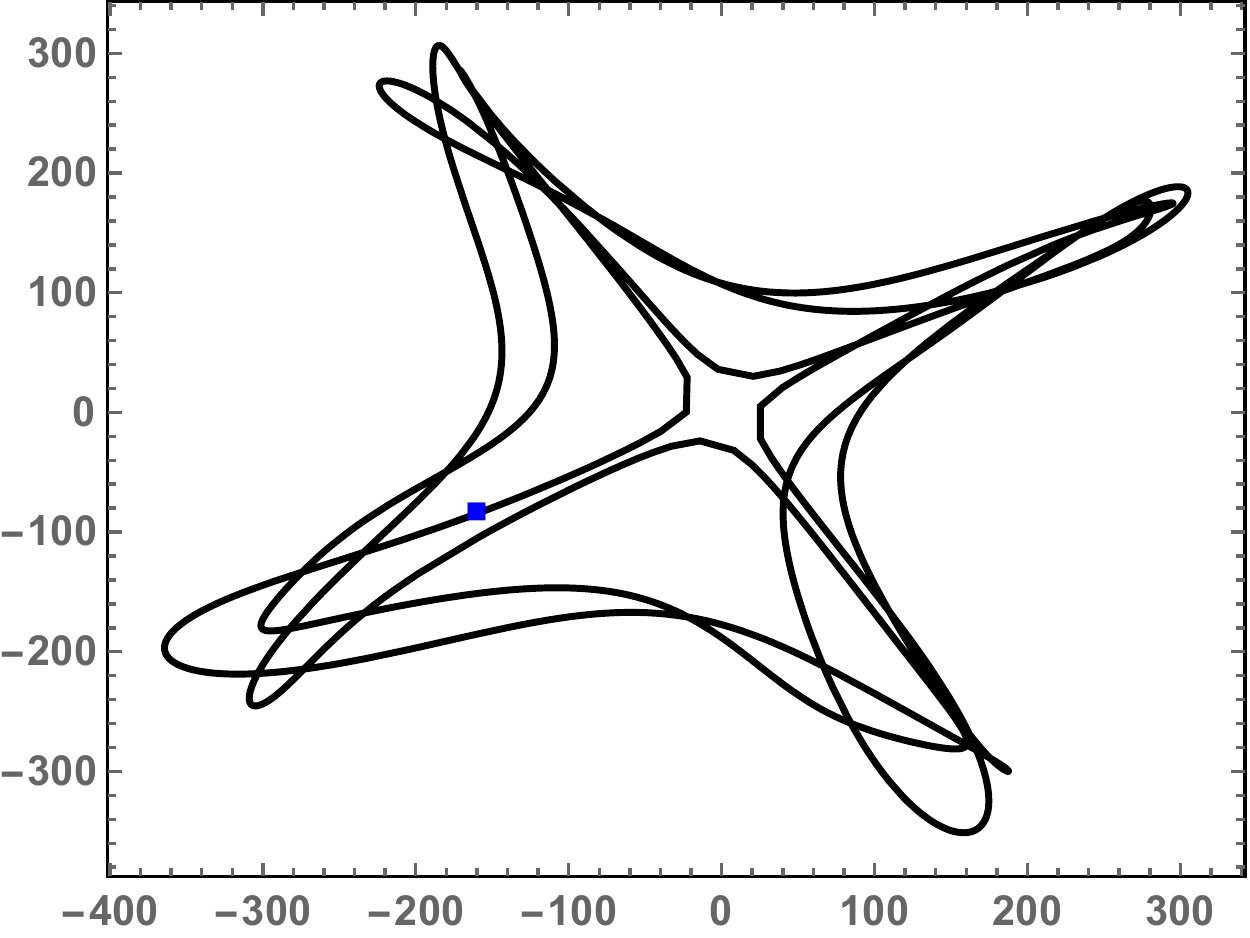}
              \caption{Initial value problem~(\ref{2Body3.1.3}),~(\ref{par:2Body3.1.3}),~(\ref{InitCond:2Body3.1.3}). Trajectory, in the complex $x$-plane, of  $x_2(t)$; 
              period $24$. The   square indicates the initial condition $x_2(0)=-160.42 - 84.73 \;\mathbf{i}$.}
              \label{Ex313F4}
          \end{figure}
      \end{minipage}
  \end{minipage}
\smallskip

\noindent \textbf{Example 3.1.4.} In this example, the  generating model is
\begin{eqnarray}
&&\ddot{y}_1=-r^2\, \omega^2 y_1,\notag\\
&&\ddot{y}_2=-a\, \dot{y}_2,
\label{Model3.1.4}
\end{eqnarray}
where $\omega$ is a nonvanishing real number, $r$ is a nonvanishing rational number and $a$ is a positive real number.
System~\eqref{Model3.1.3} is Hamiltonian and integrable; its solution 
\begin{eqnarray}
&&y_1(t)=y_1(0)\cos(r\, \omega\, t)+\frac{1}{r\, \omega}\dot{y}_1(0) \sin(r\, \omega\, t),\notag\\
&&y_2(t)=y_2(0)+\frac{1}{a} \, \dot{y}_2(0)\left[1-\exp(-at) \right]
\end{eqnarray}
 is asymptotically isochronous.

Via Theorem~\ref{Theorem2ndOrder}, model~\eqref{Model3.1.4} generates the following solvable $2$-body problem, see~\eqref{2BodyGeneral}:
\begin{eqnarray}
&&\ddot{x}_1= \frac{\dot{x}_1(a\, x_2+m_1\, \dot{x}_1+2\, \dot{x}_2)}{x_1-x_2}-r^2\,\omega^2\, \frac{x_1\left[(m_1+1)x_1+x_2 \right]}{x_1-x_2}\notag\\
&&+a\, \frac{x_1(m_1 \dot{x}_1+\dot{x}_2)}{x_1-x_2}
,\notag\\
&&\ddot{x}_2=-(m_1+1) \frac{\dot{x}_1 (m_1 \dot{x}_1+2\,\dot{x}_2)}{x_1-x_2}+r^2 \,\omega^2\,
\frac{(m_1 x_1+x_2)\left[(m_1+1) x_1+x_2 \right]}{x_1-x_2}\notag\\
&&-a(m_1+1)\frac{x_1(m_1\dot{x}_1+\dot{x}_2)+x_2\dot{x}_1}{x_1-x_2}.
\label{2Body3.1.4}
\end{eqnarray}
System~\eqref{2Body3.1.4} is asymptotically isochronous because model~\eqref{Model3.1.4} is asymptotically isochronous, see the reasoning below display~\eqref{2Body3.1.1}.

In Figures~\ref{Ex314F1},~\ref{Ex314F2},~\ref{Ex314F3},~\ref{Ex314F4} we provide the plots of the solutions of system~(\ref{2Body3.1.4}) with the parameters
\begin{equation}
m_1=6,\;\; r=\frac{1}{3},\; \;\omega=2\pi,\;\; a=0.1, 
\label{par:2Body3.1.4}
\end{equation}
satisfying the initial conditions
\begin{eqnarray}
&&x_1(0) =295.50 + 156.68 \; \mathbf{i}, \;\;\;\;\;\;\;x_1'(0) =14.47 + 5.64 \; \mathbf{i}, \notag\\
&&x_2(0) =-1082.47 - 679.55 \; \mathbf{i}, \;\;x_2'(0) =0.36 + 1.79 \; \mathbf{i}.
\label{InitCond:2Body3.1.4}
\end{eqnarray}

\begin{minipage}{\linewidth}
      \centering
      \begin{minipage}{0.45\linewidth}
          \begin{figure}[H]
              \includegraphics[width=\linewidth]{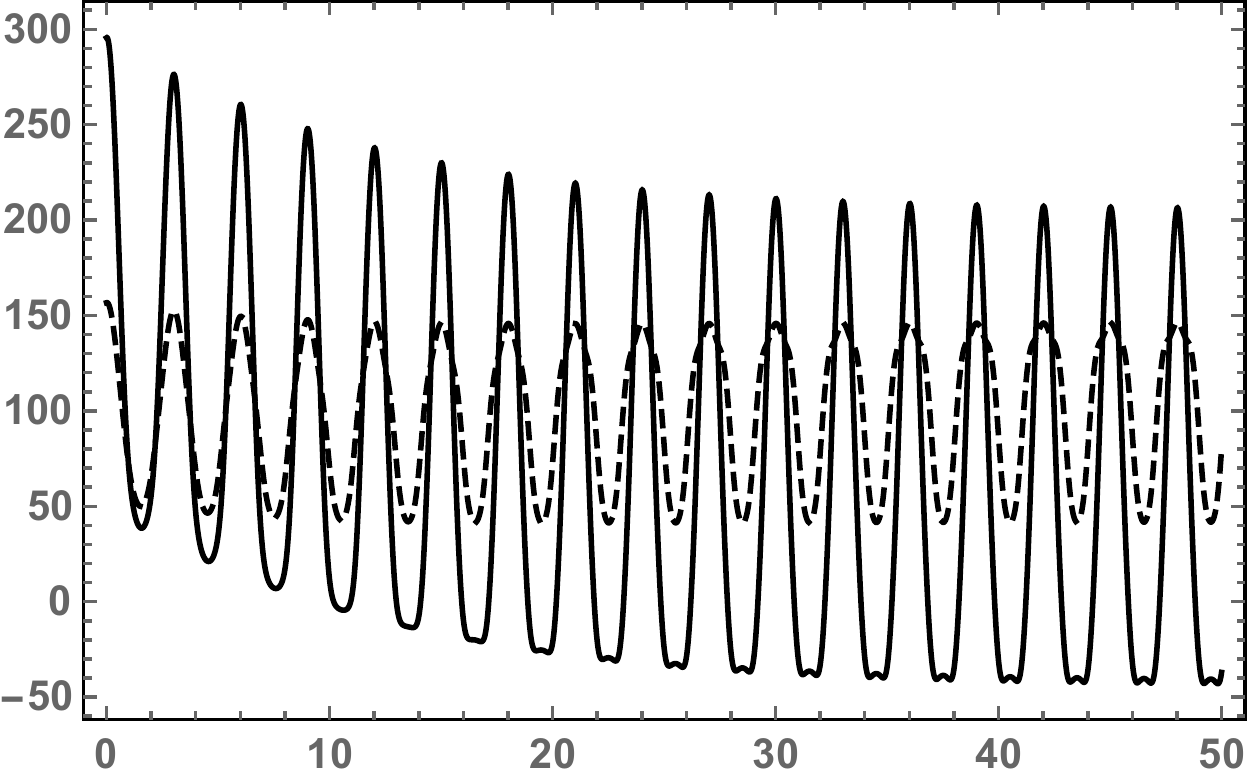}
              \caption{Initial value problem~(\ref{2Body3.1.4}),~(\ref{par:2Body3.1.4}),~(\ref{InitCond:2Body3.1.4}). Graphs of the real (bold curve) and imaginary 
              (dashed curve) parts of the coordinate $x_1(t)$.}
              \label{Ex314F1}
          \end{figure}
      \end{minipage}
      \hspace{0.05\linewidth}
      \begin{minipage}{0.45\linewidth}
          \begin{figure}[H]
              \includegraphics[width=\linewidth]{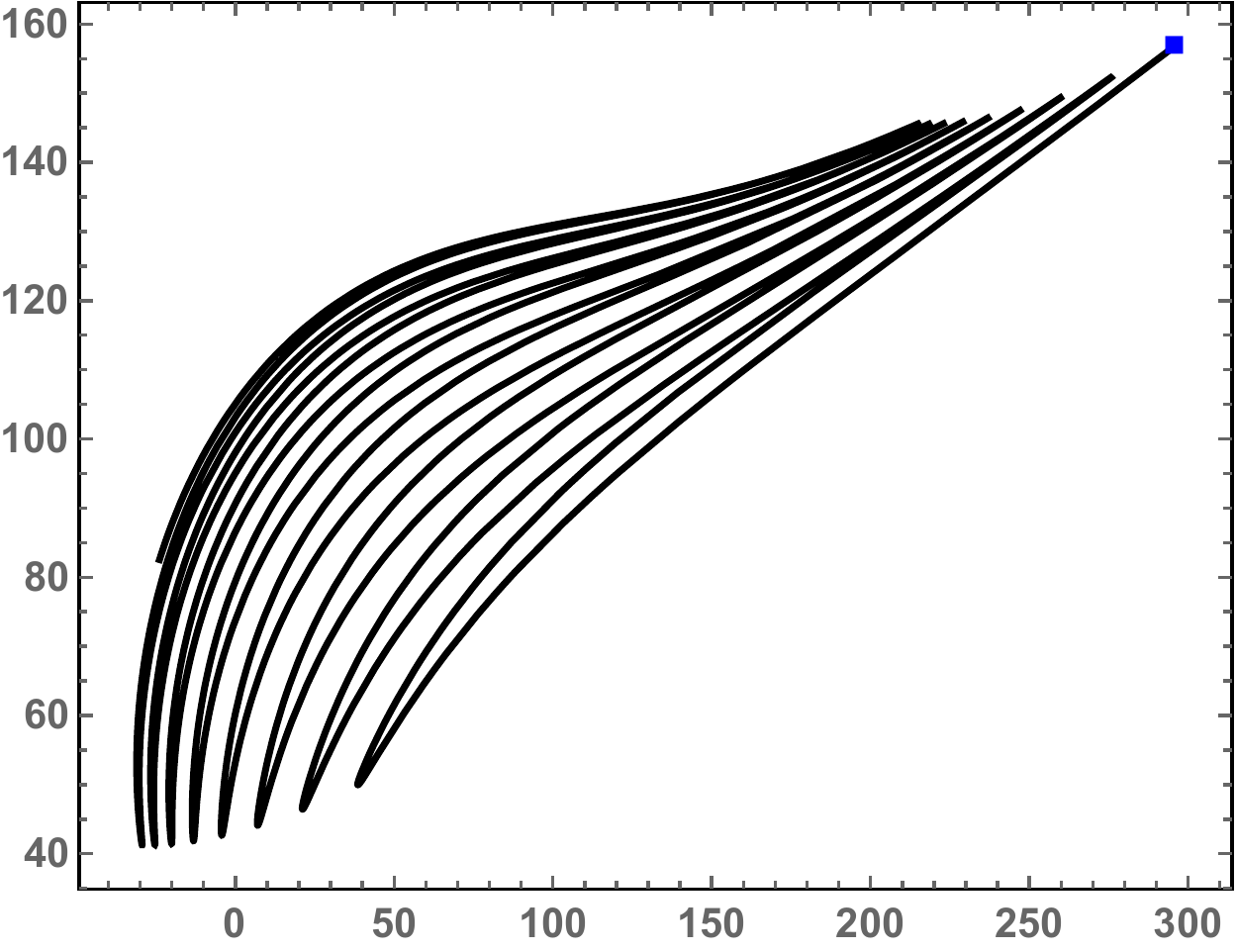}
              \caption{Initial value problem~(\ref{2Body3.1.4}),~(\ref{par:2Body3.1.4}),~(\ref{InitCond:2Body3.1.4}). Trajectory, in the complex $x$-plane, of  $x_1(t)$. The   square indicates the initial condition $x_1(0)=295.50 + 156.68 \; \mathbf{i}$.}
              \label{Ex314F2}
          \end{figure}
      \end{minipage}
  \end{minipage}
  
  \begin{minipage}{\linewidth}
      \centering
      \begin{minipage}{0.45\linewidth}
          \begin{figure}[H]
              \includegraphics[width=\linewidth]{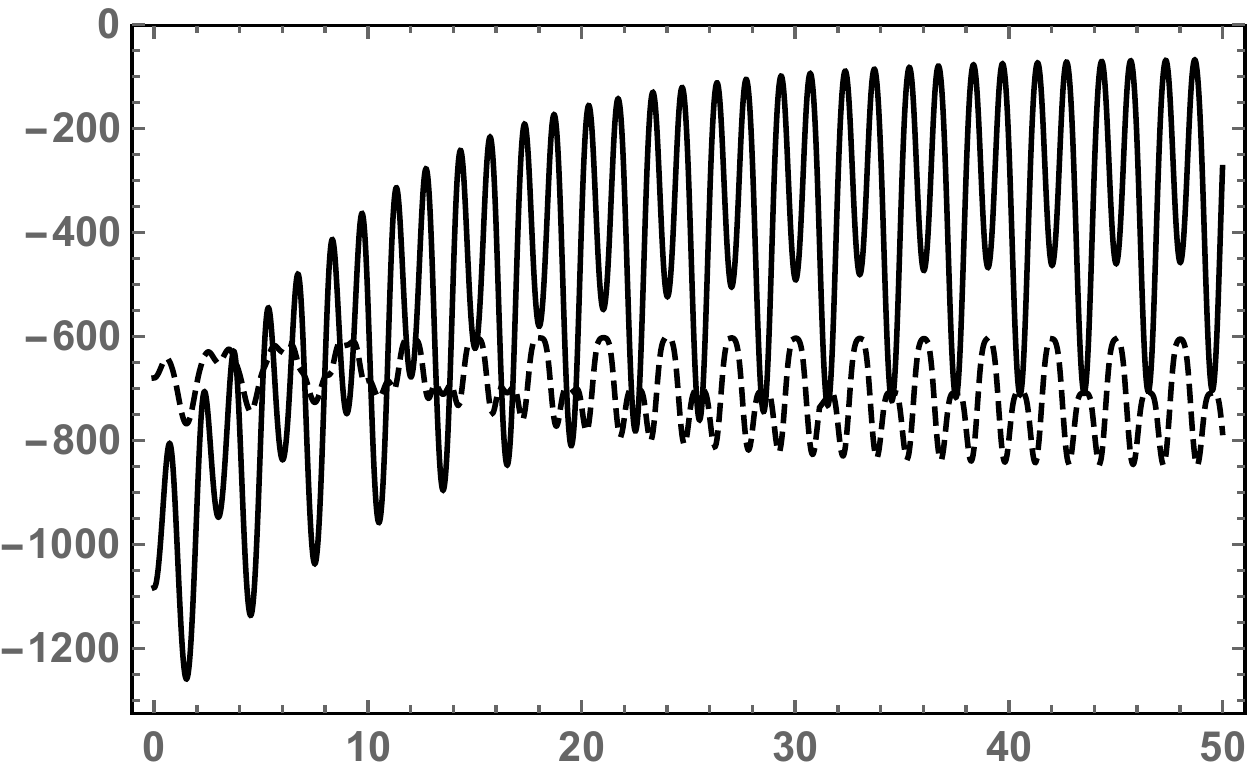}
              \caption{Initial value problem~(\ref{2Body3.1.4}),~(\ref{par:2Body3.1.4}),~(\ref{InitCond:2Body3.1.4}). Graphs of the real (bold curve) and imaginary 
              (dashed curve) parts of the coordinate $x_2(t)$.}
              \label{Ex314F3}
          \end{figure}
      \end{minipage}
      \hspace{0.05\linewidth}
      \begin{minipage}{0.45\linewidth}
          \begin{figure}[H]
              \includegraphics[width=\linewidth]{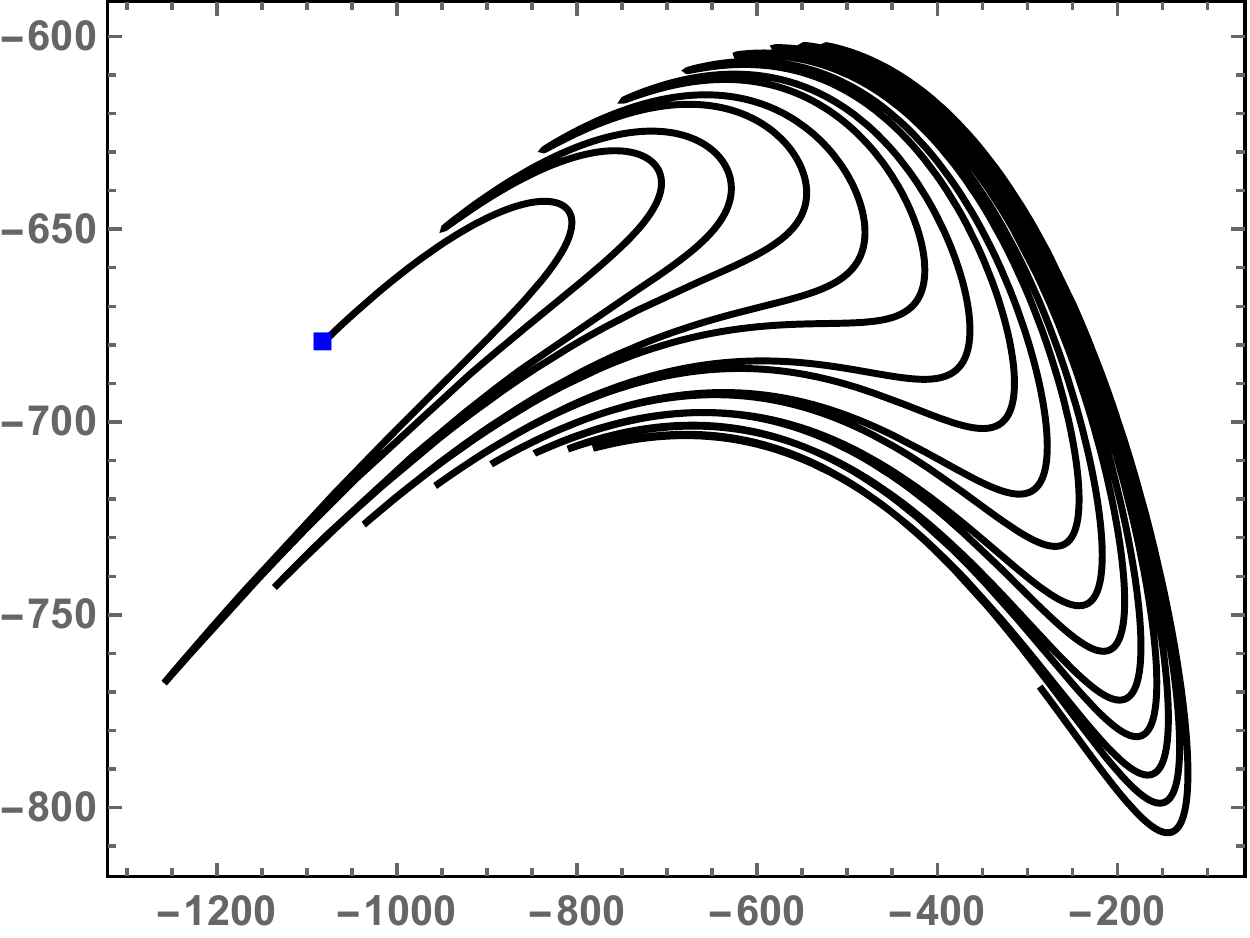}
              \caption{Initial value problem~(\ref{2Body3.1.4}),~(\ref{par:2Body3.1.4}),~(\ref{InitCond:2Body3.1.4}). Trajectory, in the complex $x$-plane, of  $x_2(t)$. The   square indicates the initial condition $x_2(0)=-1082.47 - 679.55 \; \mathbf{i}$.}
              \label{Ex314F4}
          \end{figure}
      \end{minipage}
  \end{minipage}
\smallskip

\subsection{Three-body problems}
 Consider the solvable $N$-body problem of Theorem~\ref{Theorem2ndOrder} with $N=3$. In this case, polynomial~\eqref{MainPoly} reduces to
\begin{eqnarray}
p_{3+m_1}(z;t)=(z-x_1)^{m_1+1}(z-x_2)(z-x_3)=z^{m_1+3}+\sum_{n=1}^{m_1+3} y_n z^{m_1+3-n}
\label{MainPolyEx3Body}
\end{eqnarray}
so that
\begin{eqnarray}
&&y_1=-(m_1+1) x_1-x_2-x_3, \notag\\
&&y_2=\binom{m_1+1}{2} x_1^2+(m_1+1)x_1(x_2+x_3)+x_2 x_3,
\notag\\
&&
y_3=-(m_1+1) x_1 x_2 x_3-\binom{m_1+1}{2} x_1^2(x_2+x_3)-\binom{m_1+1}{3} x_1^3
\label{y123Ex3Body}
\end{eqnarray}
and
\begin{eqnarray}
&&\dot{y}_1=-(m_1+1) \dot{x}_1-\dot{x}_2-\dot{x}_3, \notag\\
&&\dot{y}_2=\binom{m_1+1}{2} 2 x_1 \dot{x}_1+(m_1+1)[\dot{x}_1(x_2+x_3)+x_1(\dot{x}_2+\dot{x}_3)]+\dot{x}_2 x_3+x_2 \dot{x}_3,\notag\\
&&\dot{y}_3=-(m_1+1)(\dot{x}_1 x_2 x_3+x_1 \dot{x}_2 x_3 + x_1 x_2 \dot{x}_3)\notag\\
&&-\binom{m_1+1}{2}\Big[2 x_1 \dot{x}_1 (x_2+x_3)+ x_1^2 (\dot{x}_2+\dot{x}_3)
\Big]-3\binom{m_1+1}{3} x_1^2 \dot{x}_1,
 \label{doty123Ex3Body}
\end{eqnarray}
see~\eqref{yjThm2},~\eqref{dotyjThm2}. As for the parameters $\vec{\gamma}$ and $\underline{\alpha}$ of system~\eqref{xnddotSolvableExplicit}, we will need the following coefficients:
\begin{eqnarray*}
&&\alpha_{21}=m_1,\;\alpha_{31}=\frac{m_1(m_1+1)}{2},\; \alpha_{32}=m_1,\\
&&\gamma_1=-m_1, \; \gamma_2=-\frac{m_1(m_1+1)}{2},\; \gamma_3=-\frac{m_1(m_1^2+3 m_1+2)}{6},
\end{eqnarray*}
see~\eqref{alphabetanm},~\eqref{gamman}.
Thus, in the case where $N=3$ system~\eqref{xnddotSolvableExplicit} reduces to
\begin{eqnarray}
&&\ddot{x}_1=\dot{x}_1 \left(\frac{m_1 \dot{x}_1 + 2 \dot{x}_2}{x_1 - x_2} + 
\frac{m_1 \dot{x}_1 + 2 \dot{x}_3}{ x_1 - x_3}\right) \notag\\
&&- \frac{1}{2 (m_1 + 1) (x_1 - x_2) (x_1 - x_3)}\Big[(m_1 + 2) (m_1 + 1) x_1^2 f_1^{(2)}(\vec{y}^{(3)}, \dot{\vec{y}}^{(3)})\notag\\
&&+ 
  2 (m_1 + 1) x_1 f_2^{(2)} (\vec{y}^{(3)}, \dot{\vec{y}}^{(3)})+ 2 f_3^{(2)}(\vec{y}^{(3)}, \dot{\vec{y}}^{(3)}) \Big]
  ,\notag\\
&&\ddot{x}_2=\frac{1}{2 (x_1 - x_2) (x_2 - x_3)} 
\Big\{2 m_1 (m_1 + 1) \dot{x}_1^2 (x_3 - x_1) \notag\\
   &&+ 
   4 (m_1 + 1) \dot{x}_1 \dot{x}_2 (x_3 - x_2) + 
   4 \dot{x}_2 \dot{x}_3 (x_1 - x_2) \notag\\
   &&+ \big[m_1 (m_1 + 1) x_1^2 + 
      2  x_2 (m_1 x_1 + x_2) \big] f_1^{(2)} (\vec{y}^{(3)}, \dot{\vec{y}}^{(3)}) \notag\\
      &&+ 2 ( m_1 x_1 + x_2) f_2^{(2)} (\vec{y}^{(3)}, \dot{\vec{y}}^{(3)}) + 2 f_3^{(2)} (\vec{y}^{(3)}, \dot{\vec{y}}^{(3)})\Big\}
,\notag\\
&&\ddot{x}_3=-\frac{1}{2 (x_1 - x_3) (x_2 - x_3)} \Big\{2 m_1 (m_1 + 1)  \dot{x}_1^2 (x_2 - x_1) \notag\\
&&+ 
   4 (m_1 + 1) \dot{x}_1 \dot{x}_3 (x_2 - x_3) + 4 \dot{x}_2 \dot{x}_3 (x_1 - x_3) \notag\\
   &&+
   \big[m_1 (m_1 + 1) x_1^2 + 2 x_3 (m_1 x_1 + x_3)\big] f_1^{(2)}(\vec{y}^{(3)}, \dot{\vec{y}}^{(3)})  \notag\\
   &&+ 
   2 (m_1 x_1 + x_3) f_2^{(2)}(\vec{y}^{(3)}, \dot{\vec{y}}^{(3)})  + 2 f_3^{(2)}(\vec{y}^{(3)}, \dot{\vec{y}}^{(3)}) \Big\},
\label{3BodyGeneral}
\end{eqnarray}
where $\vec{y}^{(3)}=(y_1, y_2,y_3)$ and $\dot{\vec{y}}^{(3)}=(\dot{y}_1, \dot{y}_2, \dot{y}_3)$ are given by~\eqref{y123Ex3Body} and~\eqref{doty123Ex3Body}. By Theorem~\ref{Theorem2ndOrder}, the last $3$-body problem is algebraically solvable if system~\eqref{yddot_SystemThm} with $N=3$ is algebraically solvable.

\noindent \textbf{Example 3.2.1.} In this example, the  generating model is
\begin{eqnarray}
\ddot{y}_m=\mathbf{i} \, r_m \, \omega\, \dot{y}_m, \;\;\;m=1,2,3,
\label{Model3.2.1}
\end{eqnarray}
where $\omega$ is a nonvanishing real number and $r_1, r_2, r_3$ are nonvanishing rational numbers.
System~\eqref{Model3.2.1} is Hamiltonian and integrable; its solution~\eqref{ySolnModel3.1.1} with $m=1,2,3$
 is isochronous with a period that is an integer multiple of $2\pi/|\omega|$.

Via Theorem~\ref{Theorem2ndOrder}, model~\eqref{Model3.2.1} generates the following solvable $3$-body problem, see~\eqref{3BodyGeneral}:

\begin{subequations}
\begin{eqnarray}
&&\ddot{x}_1= \frac{\mathbf{i} }{2 (x_1 -
      x_2) (x_1 - x_3)}\notag\\
&&\cdot \Bigg\{\Big[(1+m_1)(2+  m_1) r_1 - 
        m_1 (2 (1 + m_1) r_2 + (1 - m_1) r_3)\Big] \omega x_1^2 
        \dot{x}_1\notag\\
         &&+ \Big[(2 + m_1) r_1 - 2 (1 + m_1) r_2 + m_1 r_3\Big] \omega x_1^2 (\dot{x}_2 +\dot{x}_3)\notag\\
        &&- 
     2 \Big[r_2 + m_1 (r_2 -  r_3)\Big] \omega x_1 \dot{x}_1(x_2+x_3)\notag\\
     &&
    + 2 r_3 \omega x_2 x_3 \dot{x}_1
    +2 \mathbf{i} m_1 \dot{x}_1^2(-2 x_1+x_2+x_3)
   \notag\\
     &&+ 
     2 (-r_2 + r_3) \omega x_1 (x_3 \dot{x}_2 +x_2 \dot{x}_3)+ 
     4 \mathbf{i}  \dot{x}_1 \dot{x}_2 (-x_1+x_3) \notag\\
     &&+
     4 \mathbf{i}  \dot{x}_1 \dot{x}_3 (-x_1+x_2) \Bigg\},
\end{eqnarray}
\begin{eqnarray}
&&\ddot{x}_2=\frac{1}{2 (x_1 - 
     x_2) (x_2 - x_3)}\notag\\
   &&\cdot\Bigg\{-\mathbf{i} m_1 (1 + m_1) \Big[r_1 (1+ m_1) - 2 m_1 r_2+( -1+m_1) r_3 \Big] \omega x_1^2 \dot{x}_1\notag\\
    && - 
   \mathbf{i} m_1 (1 + m_1) (r_1 - 2 r_2 + r_3) \omega x_1\Big[x_1 (\dot{x}_2+\dot{x}_3)+2 x_2 \dot{x}_1 \Big]\notag\\
     && +
   2 \mathbf{i} (1 + m_1) \omega \dot{x}_1 \Big[ -(r_1 - r_2) x_2^2 +  (r_2 - r_3)  x_3 (m_1 x_1+x_2) \Big]\notag\\
   &&+ 2 m_1 (1 + m_1)  \dot{x}_1^2(-x_1+x_3) -2 \mathbf{i} \Big[m_1 (r_1 - r_2) - r_2\Big] \omega x_1 x_2 \dot{x}_2\notag\\
   &&
   +2 \mathbf{i} \Big[m_1 (r_2 - r_3) - r_3\Big] \omega x_1 x_3 \dot{x}_2\notag\\
   &&
   + 2 \mathbf{i}  \omega x_2 \dot{x}_2(-r_1 x_2+r_2 x_3)+ 
   4 (1 + m_1)  \dot{x}_1 \dot{x}_2(-x_2+x_3) \notag\\
   && - 
   2 \mathbf{i} \Big[-r_2 + r_3 + m_1 (r_1 - 2 r_2 + r_3)\Big] \omega x_1 x_2 \dot{x}_3 \notag\\
   && - 
   2 \mathbf{i} (r_1 - r_2) \omega x_2^2 \dot{x}_3 + 
   4 \dot{x}_2 \dot{x}_3 (x_1 -x_2)\Bigg\},
\end{eqnarray}
\begin{eqnarray}
&&\ddot{x}_3=\frac{1}{2 (x_1 - 
     x_3) (x_2 - x_3)}\notag\\
   && \cdot
\Bigg\{\mathbf{i} m_1 (1 + m_1) \Big[(1+m_1)r_1 - 2 m_1 r_2 +(-1+m_1) r_3\Big] \omega x_1^2 \dot{x}_1\notag\\
 && + 
   \mathbf{i} m_1 (1 + m_1) (r_1 - 2 r_2 + r_3) \omega x_1\Big[ x_1( \dot{x}_2+\dot{x}_3) +2 x_3 \dot{x}_1\Big]\notag\\
   &&  - 
   2 \mathbf{i}  (1 + m_1) (r_2 - r_3) \omega  x_2 \dot{x}_1 (m_1 x_1+x_3) \notag\\
   && + 
   2 \mathbf{i} (r_1 - r_2) \omega x_3^2\Big[  (1 + m_1) \dot{x}_1+\dot{x}_2\Big] \notag\\
   && + 
   2 m_1 (1 + m_1)  \dot{x}_1^2 (x_1-x_2)  \notag\\
   && + 2 \mathbf{i} \Big[-r_2 + r_3 + m_1 (r_1 - 2 r_2 + r_3)\Big] \omega x_1 x_3 \dot{x}_2  \notag\\
   && - 2 \mathbf{i} \Big[m_1 (r_2 - r_3) - r_3\Big] \omega x_1 x_2 
   \dot{x}_3 
   + 
   2 \mathbf{i} \Big[m_1 (r_1 - r_2) - r_2\Big] \omega x_1 x_3 \dot{x}_3 \notag\\
   &&+ 2 \mathbf{i}  \omega  x_3 \dot{x}_3(-r_2 x_2+r_1 x_3)  + 
   4 (1 + m_1) \dot{x}_1 \dot{x}_3(-x_2 +x_3)  \notag\\
   &&
 +
   4  \dot{x}_2 \dot{x}_3 (-x_1+ x_3)\Bigg\}.
\end{eqnarray}
\label{3Body3.2.1}
\end{subequations}
System~\eqref{3Body3.2.1} is isochronous for the same reasons that system~\eqref{2Body3.1.1} is isochronous, see the paragraph following display~\eqref{2Body3.1.1}.

In Figures~\ref{Ex321F1},~\ref{Ex321F2},~\ref{Ex321F3},~\ref{Ex321F4},~\ref{Ex321F5},~\ref{Ex321F6} we provide the plots of the solutions of system~(\ref{3Body3.2.1}) with the parameters
\begin{equation}
m_1=5;\;\; r_1=\frac{1}{2},\; \; r_2=\frac{1}{3}, \; \; r_3=\frac{1}{2}, \;\;\omega=2\pi, 
\label{par:3Body3.2.1}
\end{equation}
satisfying the initial conditions
\begin{eqnarray}
&&x_1(0) =-0.06 - 0.69  \;\mathbf{i}, \;\, \;\;\;\;\,x_1'(0) =3.94 - 0.82  \;\mathbf{i}, \notag\\
&&x_2(0) =8.51 + 40.06  \;\mathbf{i}, \;\;\; \;\; \;\;x_2'(0) =-52.50 + 13.06  \;\mathbf{i}, \notag\\
&&x_3(0) =-31.70 - 13.50   \;\mathbf{i}, \;\;\,x_3'(0) =-10.87 - 17.44  \;\mathbf{i}.
\label{InitCond:3Body3.2.1}
\end{eqnarray}

\begin{minipage}{\linewidth}
      \centering
      \begin{minipage}{0.45\linewidth}
          \begin{figure}[H]
              \includegraphics[width=\linewidth]{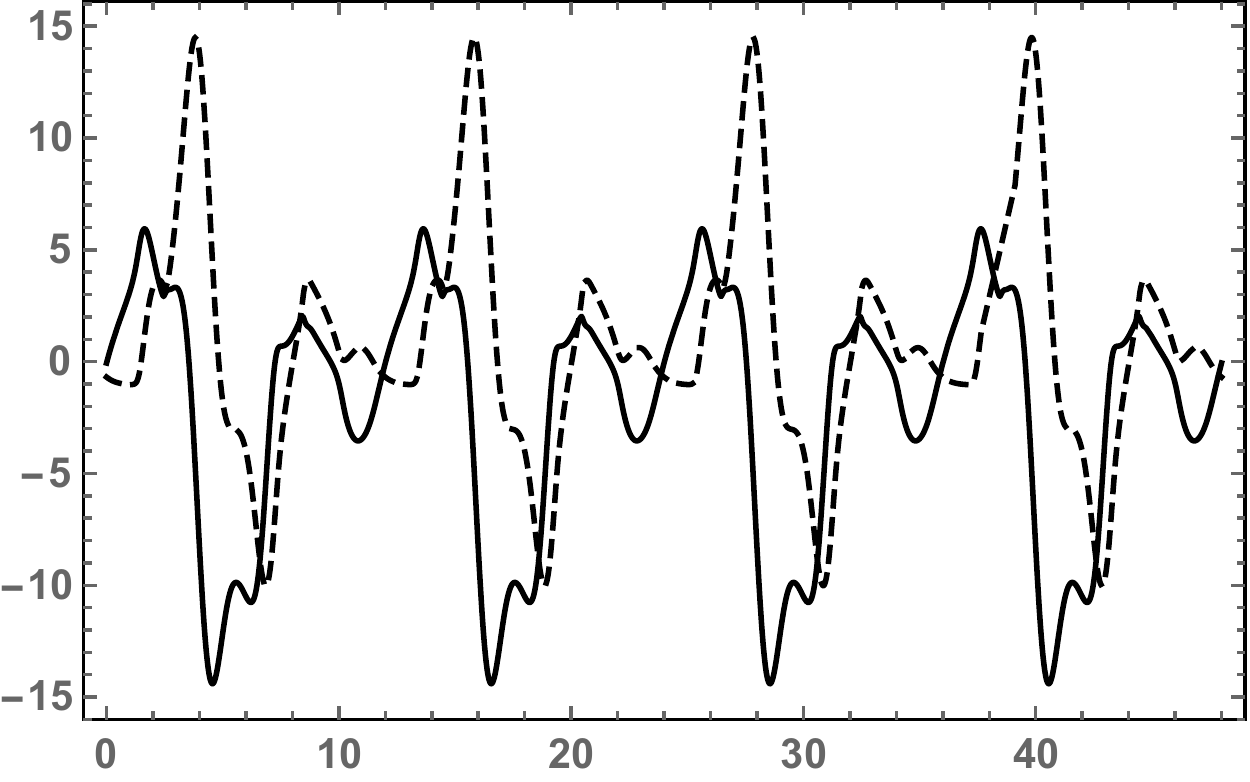}
              \caption{Initial value problem~(\ref{3Body3.2.1}),~(\ref{par:3Body3.2.1}),~(\ref{InitCond:3Body3.2.1}). Graphs of the real (bold curve) and imaginary 
              (dashed curve) parts of the coordinate $x_1(t)$; period $12$.}
              \label{Ex321F1}
          \end{figure}
      \end{minipage}
      \hspace{0.05\linewidth}
      \begin{minipage}{0.45\linewidth}
          \begin{figure}[H]
              \includegraphics[width=\linewidth]{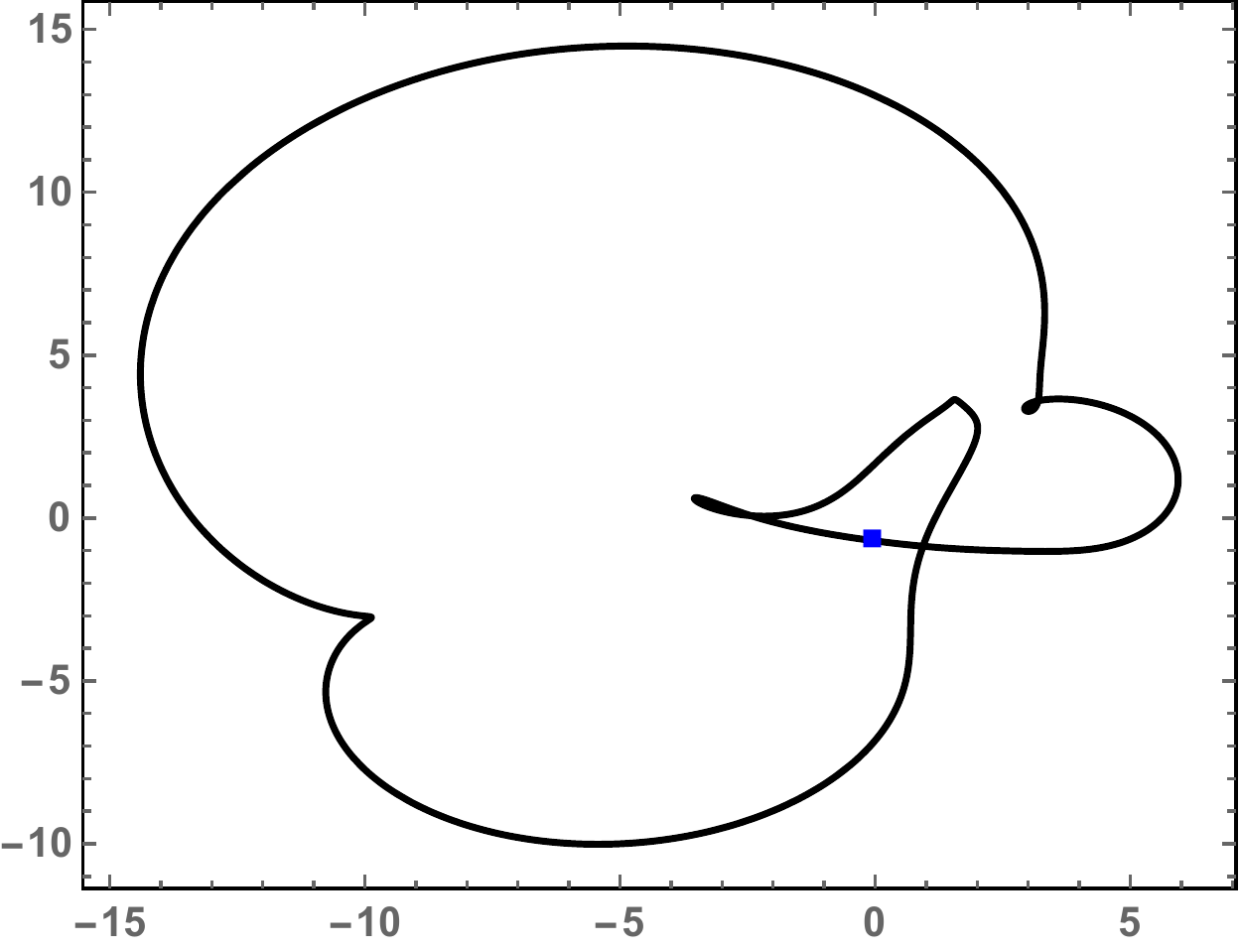}
              \caption{Initial value problem~(\ref{3Body3.2.1}),~(\ref{par:3Body3.2.1}),~(\ref{InitCond:3Body3.2.1}). Trajectory, in the complex $x$-plane, of  $x_1(t)$; 
              period $12$. The   square indicates the initial condition $x_1(0)=-0.06 - 0.69  \;\mathbf{i}$.}
              \label{Ex321F2}
          \end{figure}
      \end{minipage}
  \end{minipage}
  
  \begin{minipage}{\linewidth}
      \centering
      \begin{minipage}{0.45\linewidth}
          \begin{figure}[H]
              \includegraphics[width=\linewidth]{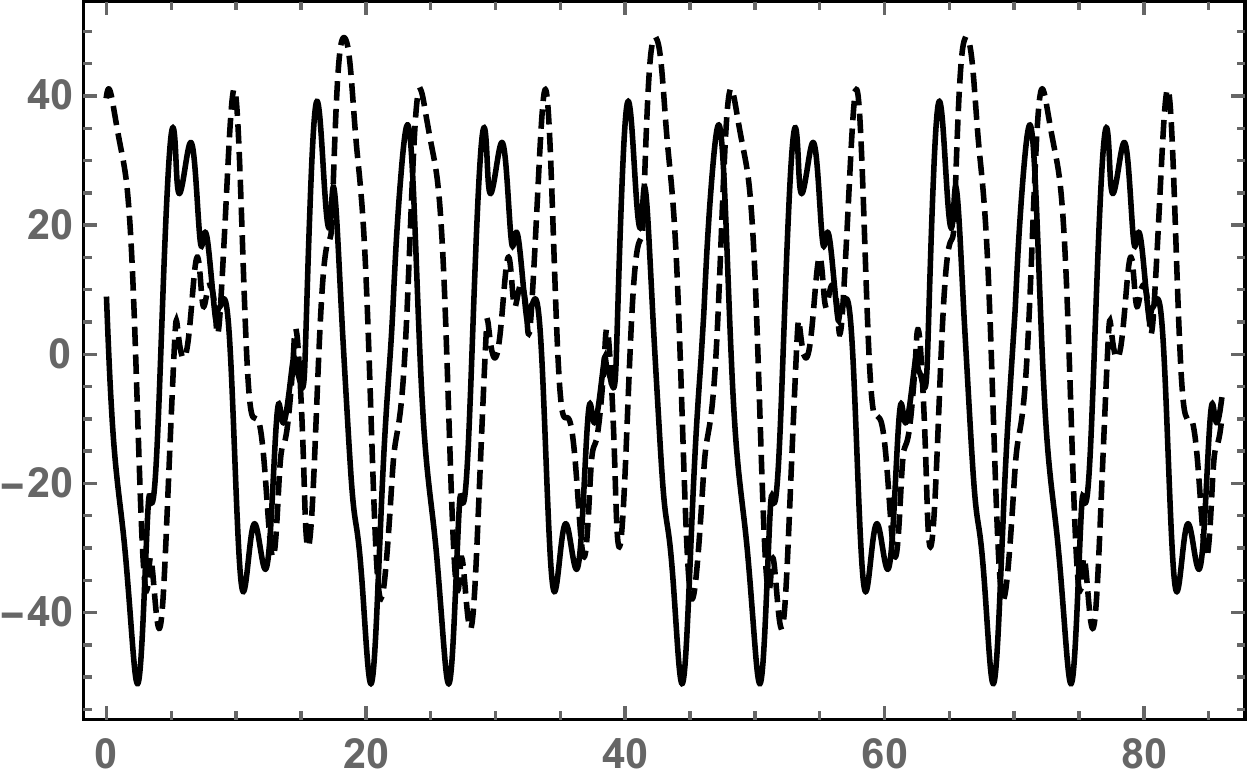}
              \caption{Initial value problem~(\ref{3Body3.2.1}),~(\ref{par:3Body3.2.1}),~(\ref{InitCond:3Body3.2.1}). Graphs of the real (bold curve) and imaginary 
              (dashed curve) parts of the coordinate $x_2(t)$; period $24$.}
              \label{Ex321F3}
          \end{figure}
      \end{minipage}
      \hspace{0.05\linewidth}
      \begin{minipage}{0.45\linewidth}
          \begin{figure}[H]
              \includegraphics[width=\linewidth]{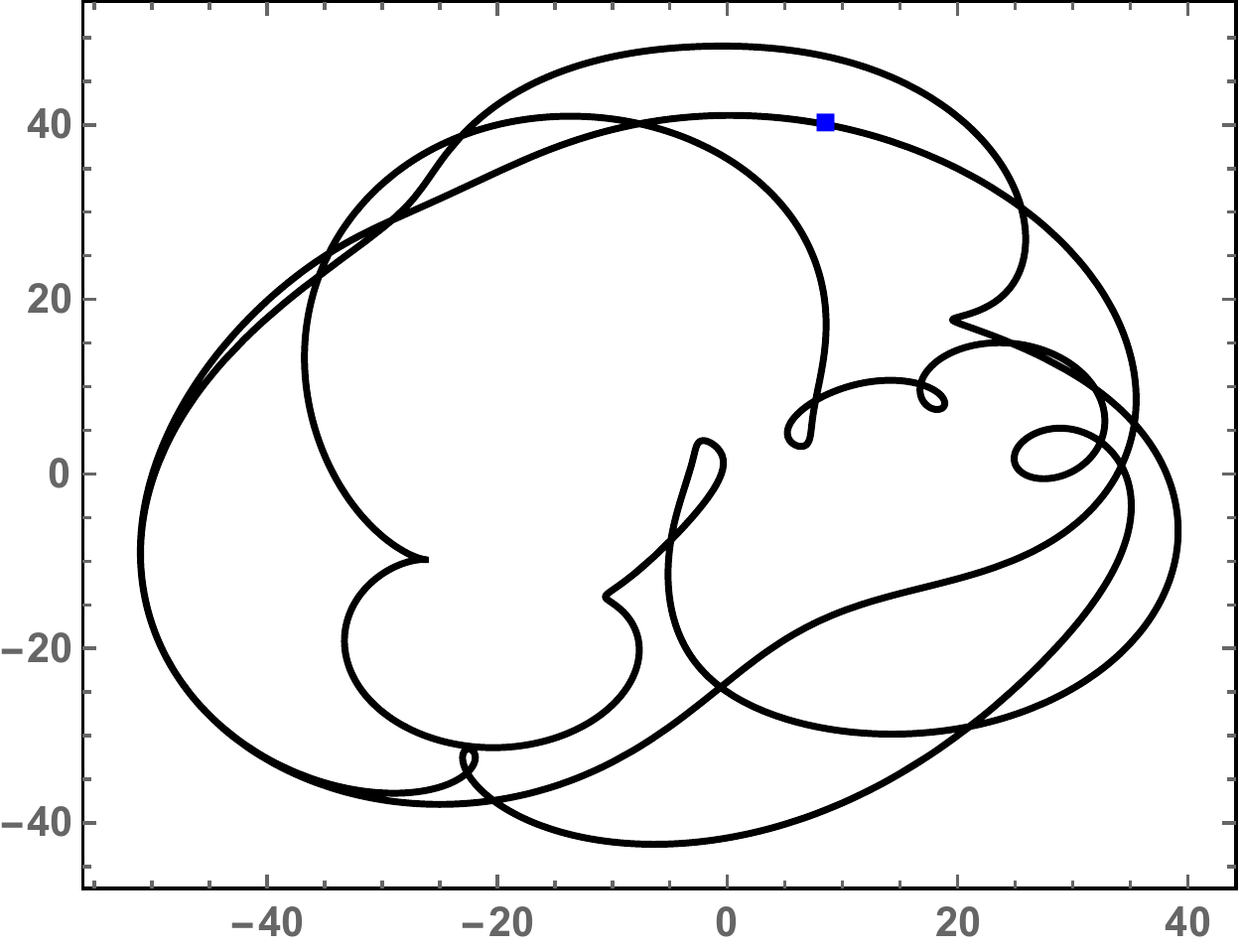}
              \caption{Initial value problem~(\ref{3Body3.2.1}),~(\ref{par:3Body3.2.1}),~(\ref{InitCond:3Body3.2.1}). Trajectory, in the complex $x$-plane, of  $x_2(t)$; 
              period $24$. The   square indicates the initial condition $x_2(0)=8.51 + 40.06  \;\mathbf{i}$.}
              \label{Ex321F4}
          \end{figure}
      \end{minipage}
  \end{minipage}
  
   \begin{minipage}{\linewidth}
      \centering
      \begin{minipage}{0.45\linewidth}
          \begin{figure}[H]
              \includegraphics[width=\linewidth]{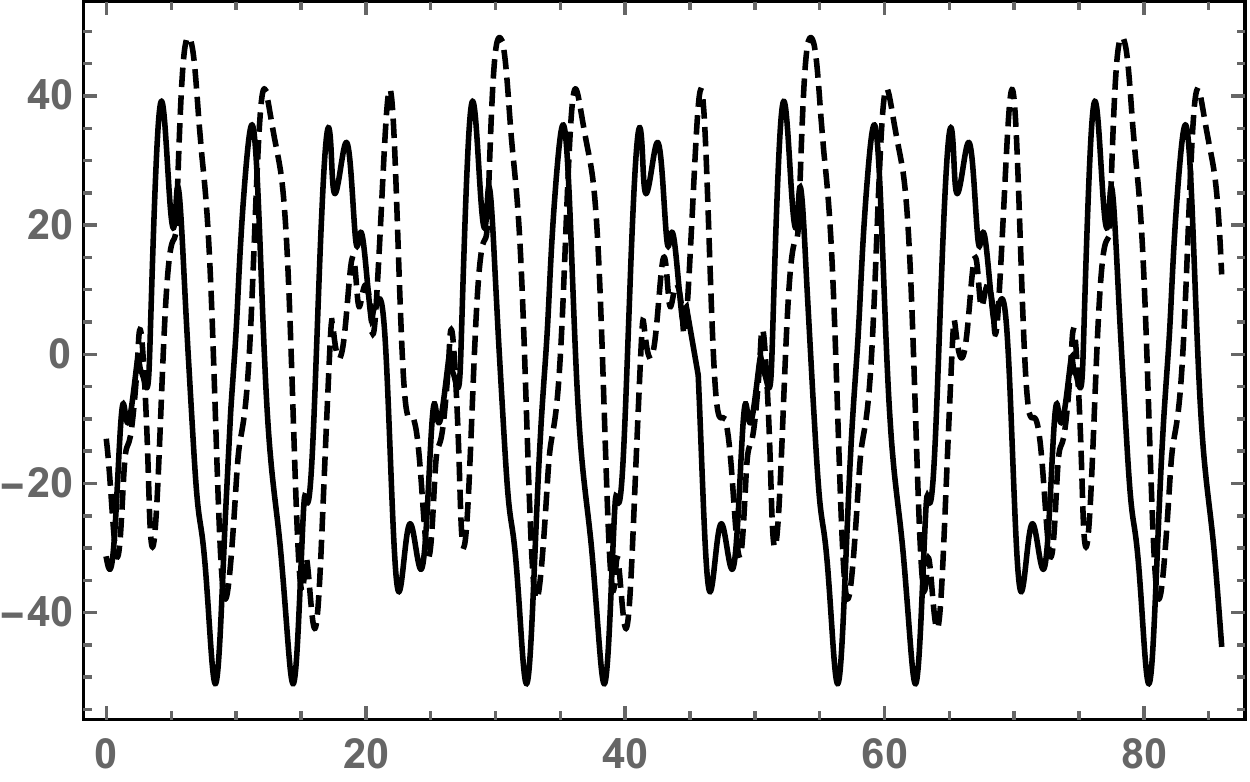}
              \caption{Initial value problem~(\ref{3Body3.2.1}),~(\ref{par:3Body3.2.1}),~(\ref{InitCond:3Body3.2.1}). Graphs of the real (bold curve) and imaginary 
              (dashed curve) parts of the coordinate $x_3(t)$; period $24$.}
              \label{Ex321F5}
          \end{figure}
      \end{minipage}
      \hspace{0.05\linewidth}
      \begin{minipage}{0.45\linewidth}
          \begin{figure}[H]
              \includegraphics[width=\linewidth]{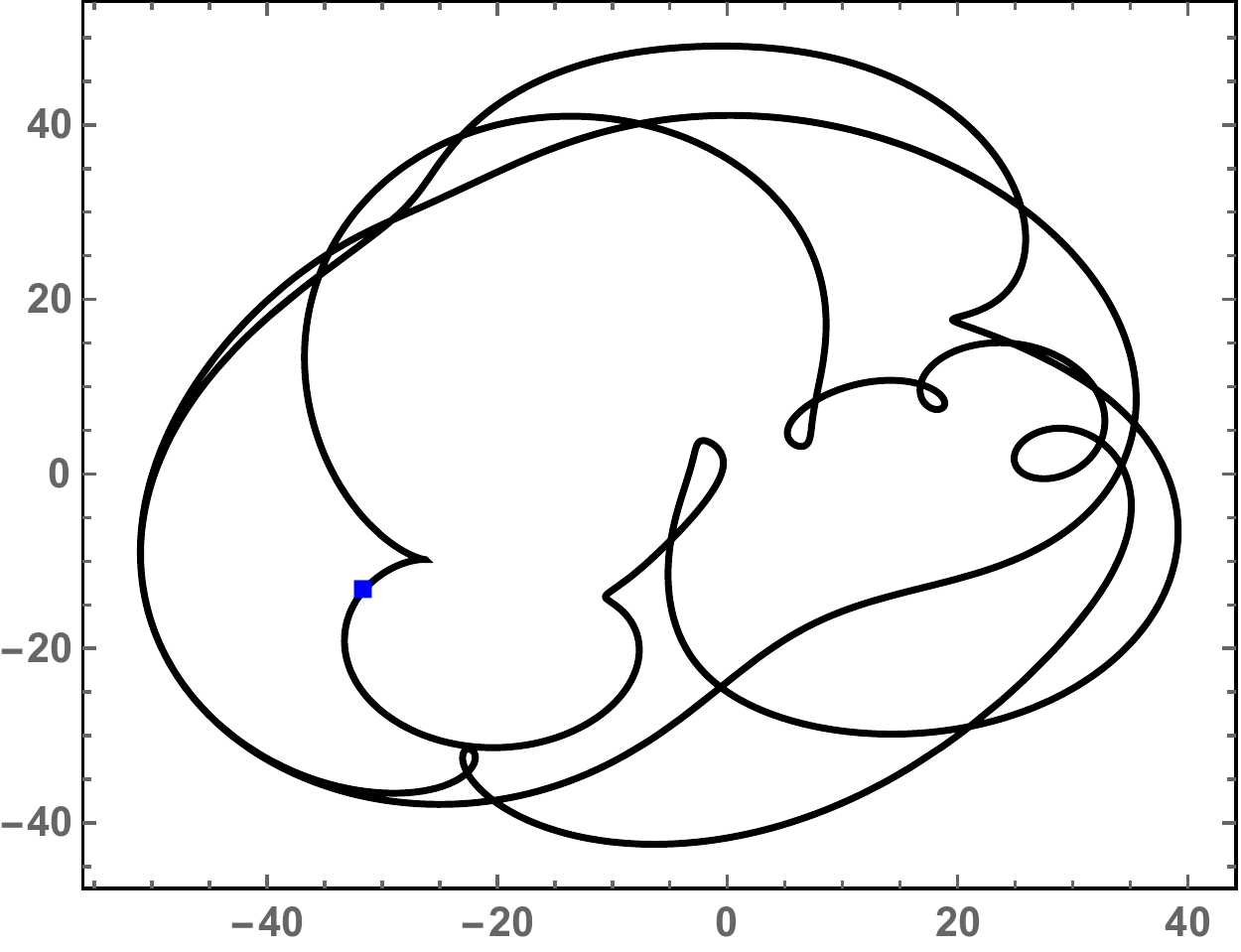}
              \caption{Initial value problem~(\ref{3Body3.2.1}),~(\ref{par:3Body3.2.1}),~(\ref{InitCond:3Body3.2.1}). Trajectory, in the complex $x$-plane, of  $x_3(t)$; 
              period $24$. The   square indicates the initial condition $x_3(0)=-31.70 - 13.50   \;\mathbf{i}$.}
              \label{Ex321F6}
          \end{figure}
      \end{minipage}
  \end{minipage}

\smallskip

\noindent \textbf{Example 3.2.2.} In this example, the  generating model is
\begin{eqnarray}
\ddot{y}_m=- r_m^2 \, \omega^2\, {y}_m, \;\;\;m=1,2,3,
\label{Model3.2.2}
\end{eqnarray}
where, as before, $\omega$ is a nonvanishing real number and $r_1, r_2, r_3$ are nonvanishing rational numbers.
System~\eqref{Model3.2.2} is Hamiltonian and integrable; its solution~\eqref{ySolnModel3.1.2} with $m=1,2,3$
 is isochronous with a period that is an integer multiple of $2\pi/|\omega|$.

Via Theorem~\ref{Theorem2ndOrder}, model~\eqref{Model3.2.2} generates the following solvable $3$-body problem, see~\eqref{3BodyGeneral}:

\begin{subequations}
\begin{eqnarray}
&&\ddot{x}_1=\frac{1}{6 (x_1 - x_2) (x_1 - x_3)}\notag\\
&&\cdot\Bigg\{\Big[-3 (1+m_1)(2+m_1) r_1^2 + 
      m_1 (3 (1 + m_1) r_2^2 - (-1 + m_1) r_3^2)\Big] \omega^2 x_1^3 \notag\\
      &&- 
   3 \Big[(2 + m_1) r_1^2 - 2 (1 + m_1) r_2^2 + m_1 r_3^2\Big] \omega^2 x_1^2 (x_2 + x_3) \notag\\
      &&+ 
   6 x_1 \Big[(r_2^2 - r_3^2) \omega^2 x_2 x_3 + 
      2 \dot{x}_1(m_1 \dot{x}_1 + \dot{x}_2 + 
         \dot{x}_3)\Big]\notag\\
         &&
         - 
   6 \dot{x}_1 \Big[x_3 (m_1 \dot{x}_1 + 2 \dot{x}_2) + 
      x_2 (m_1 \dot{x}_1 + 
         2 \dot{x}_3)\Big]\Bigg\},
\end{eqnarray}
\begin{eqnarray}
&&\ddot{x}_2=\frac{1}{2 (x_1 - x_2) (x_2 - x_3)}\notag\\
&&\cdot \Bigg\{ r_1^2 \omega^2 \Big[ m_1 (1 + m_1) x_1^2 + 2 m_1 x_1 x_2 + 
      2 x_2^2\Big] \Big[ (1 + m_1) x_1 + x_2 + x_3\Big] \notag\\
      &&+ 
   \frac{1}{3} (1 + m_1) r_3^2 \omega^2 x_1 \Big[ (-1 + m_1) m_1 x_1^2 + 6 x_2 x_3 + 
      3 m_1 x_1 (x_2 + x_3)\Big]\notag\\
      && - 
   2 r_2^2 \omega^2 (m_1 x_1 + x_2) \Big[
   \frac{1}{2} m_1 (1 + m_1) x_1^2 + 
      x_2 x_3 + (1 + m_1) x_1 (x_2 + x_3)\Big]\notag\\
      && + 
   2 m_1 (1 + m_1) (-x_1 + x_3) \dot{x}_1^2 + 
   4 (1 + m_1) (-x_2 + x_3) \dot{x}_1 \dot{x}_2 \notag\\
   &&+ 4 (x_1 - x_2) \dot{x}_2 \dot{x}_3\Bigg\},
\end{eqnarray}
\begin{eqnarray}
&&\ddot{x}_3=-\frac{1}{2 (x_1 - x_3) (x_2 - x_3)}\notag\\
&&\cdot \Bigg\{r_1^2 \omega^2 \Big[(1 + m_1) x_1 + x_2 + 
        x_3\Big] \Big[m_1 (1 + m_1) x_1^2 + 2 m_1 x_1 x_3 + 2 x_3^2\Big] \notag\\
        && + 
     \frac{1}{3} (1 + m_1) r_3^2 \omega^2 x_1 \Big[(-1 + m_1) m_1 x_1^2 + 6 x_2 x_3 + 
        3 m_1 x_1 (x_2 + x_3)\Big] \notag\\
        &&- 
     2 r_2^2 \omega^2 (m_1 x_1 + x_3) \Big[\frac{1}{2} m_1 (1 + m_1) x_1^2 + 
        x_2 x_3 + (1 + m_1) x_1 (x_2 + x_3)\Big] \notag\\
        &&+ 
     2 m_1 (1 + m_1) (-x_1 + x_2) \dot{x}_1^2 + 
     4 (1 + m_1) (x_2 - x_3) \dot{x}_1 \dot{x}_3\notag\\
     &&+ 
     4 (x_1 - x_3) \dot{x}_2 \dot{x}_3\Bigg\}.
\end{eqnarray}
\label{3Body3.2.2}
\end{subequations}
System~\eqref{3Body3.2.2} is isochronous for the same reasons that system~\eqref{2Body3.1.1} is isochronous, see the paragraph following display~\eqref{2Body3.1.1}.

In Figures~\ref{Ex322F1},~\ref{Ex322F2},~\ref{Ex322F3},~\ref{Ex322F4},~\ref{Ex322F5},~\ref{Ex322F6} we provide the plots of the solutions of system~(\ref{3Body3.2.2}) with the parameters
\begin{equation}
m_1=5;\;\; r_1=\frac{1}{2},\; \; r_2=\frac{1}{3}, \; \; r_3=\frac{1}{4}, \;\;\omega=2\pi, 
\label{par:3Body3.2.2}
\end{equation}
satisfying the initial conditions
\begin{eqnarray}
&&x_1(0) =16.92 - 28.19   \;\mathbf{i}, \;\, \;\;\;x_1'(0) =42.07 + 19.38   \;\mathbf{i}, \notag\\
&&x_2(0) =29.24 + 90.02  \;\mathbf{i}, \;\;\; \;\; \;\;x_2'(0) =-88.07 + 23.34   \;\mathbf{i}, \notag\\
&&x_3(0) =-70.22 + 40.41    \;\mathbf{i}, \;\;\,x_3'(0) =-37.49 - 99.06   \;\mathbf{i}.
\label{InitCond:3Body3.2.2}
\end{eqnarray}

\begin{minipage}{\linewidth}
      \centering
      \begin{minipage}{0.45\linewidth}
          \begin{figure}[H]
              \includegraphics[width=\linewidth]{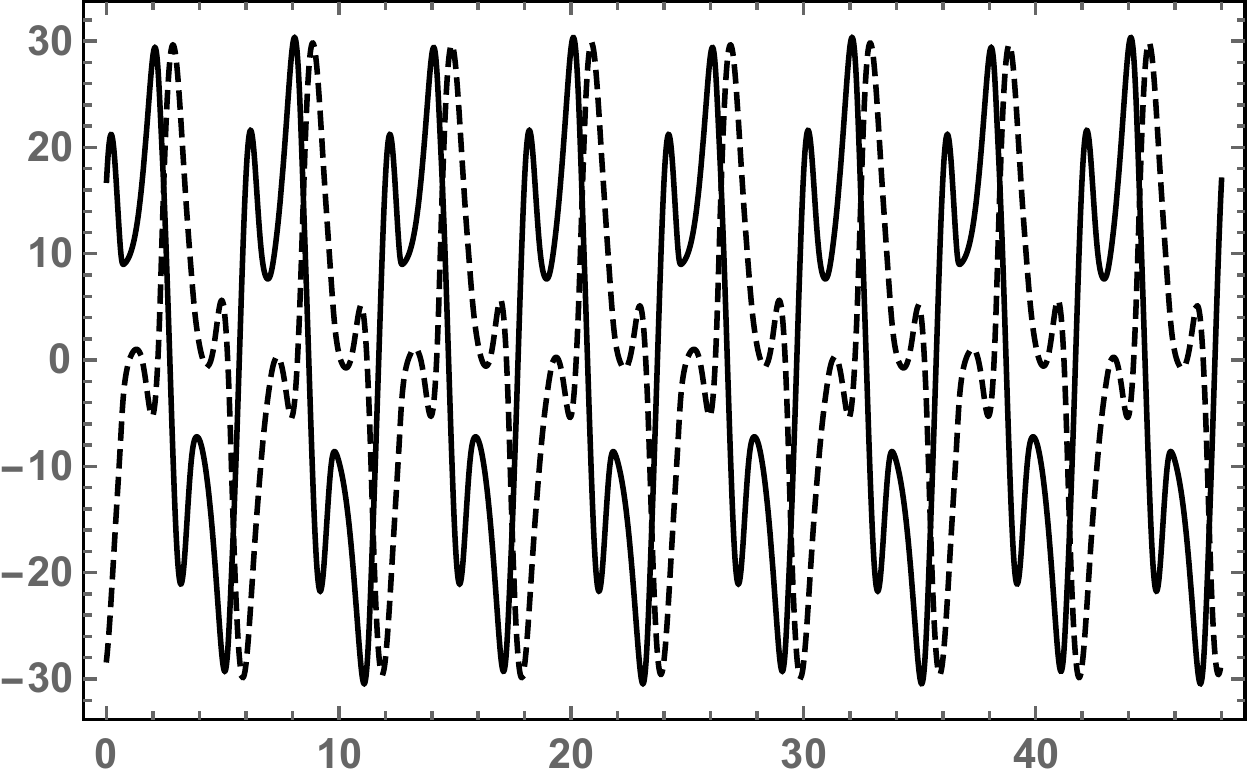}
              \caption{Initial value problem~(\ref{3Body3.2.2}),~(\ref{par:3Body3.2.2}),~(\ref{InitCond:3Body3.2.2}). Graphs of the real (bold curve) and imaginary 
              (dashed curve) parts of the coordinate $x_1(t)$; period $6$.}
              \label{Ex322F1}
          \end{figure}
      \end{minipage}
      \hspace{0.05\linewidth}
      \begin{minipage}{0.45\linewidth}
          \begin{figure}[H]
              \includegraphics[width=\linewidth]{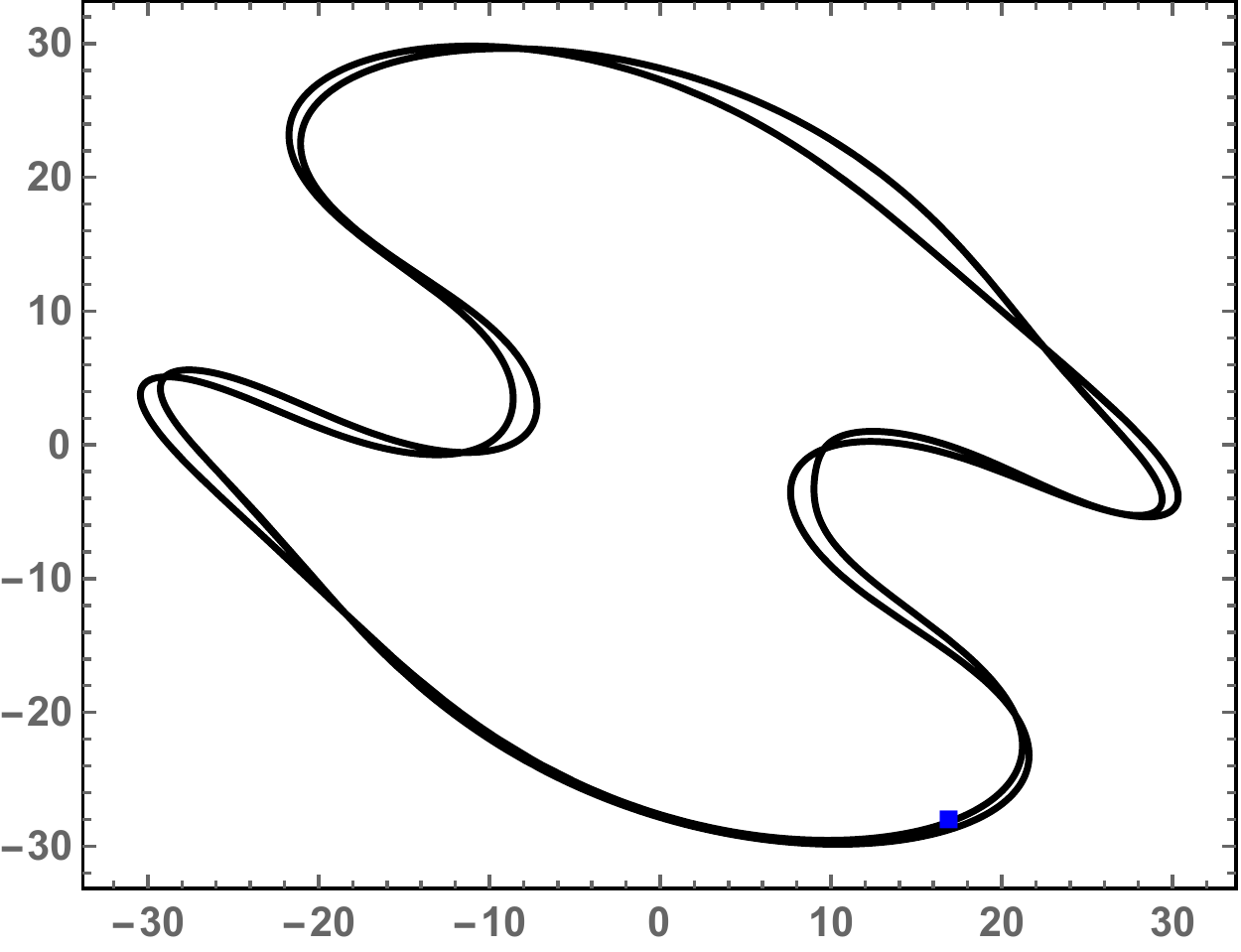}
              \caption{Initial value problem~(\ref{3Body3.2.2}),~(\ref{par:3Body3.2.2}),~(\ref{InitCond:3Body3.2.2}). Trajectory, in the complex $x$-plane, of  $x_1(t)$; 
              period $6$. The   square indicates the initial condition $x_1(0)=16.92 - 28.19   \;\mathbf{i}$.}
              \label{Ex322F2}
          \end{figure}
      \end{minipage}
  \end{minipage}
  
  \begin{minipage}{\linewidth}
      \centering
      \begin{minipage}{0.45\linewidth}
          \begin{figure}[H]
              \includegraphics[width=\linewidth]{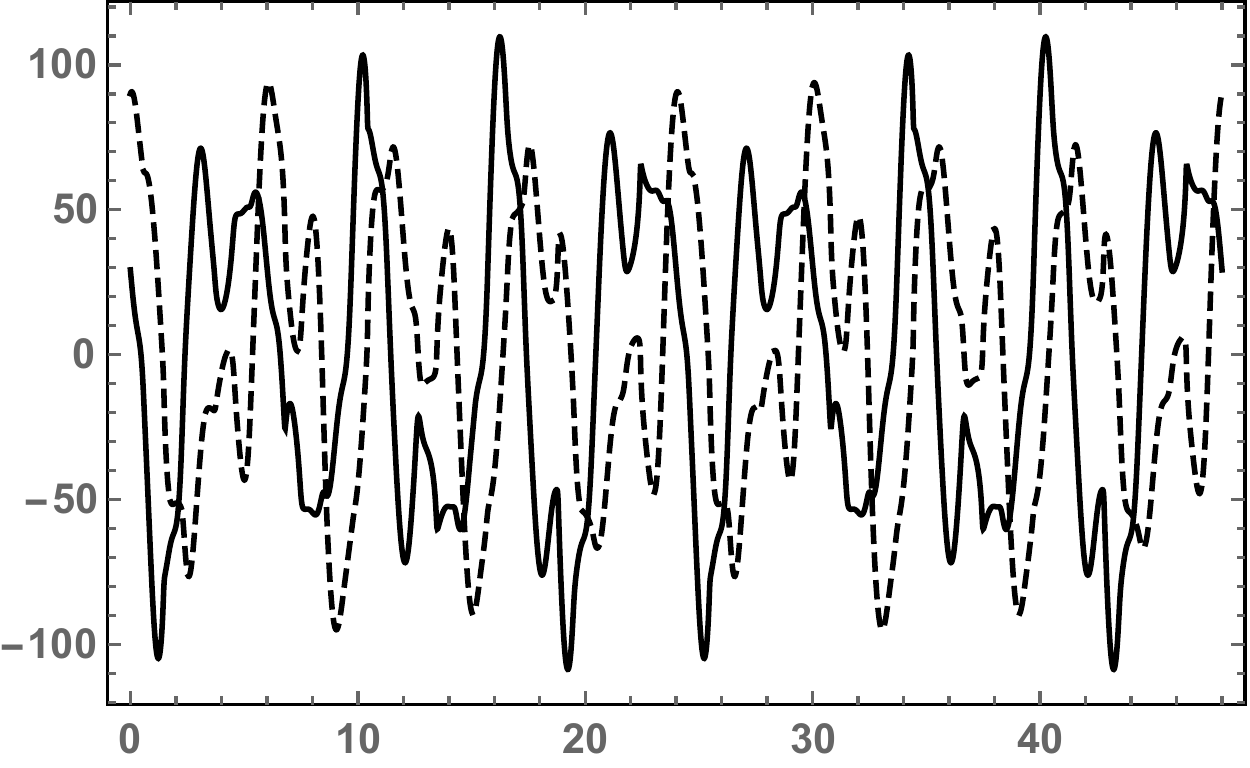}
              \caption{Initial value problem~(\ref{3Body3.2.2}),~(\ref{par:3Body3.2.2}),~(\ref{InitCond:3Body3.2.2}). Graphs of the real (bold curve) and imaginary 
              (dashed curve) parts of the coordinate $x_2(t)$; period $24$.}
              \label{Ex322F3}
          \end{figure}
      \end{minipage}
      \hspace{0.05\linewidth}
      \begin{minipage}{0.45\linewidth}
          \begin{figure}[H]
              \includegraphics[width=\linewidth]{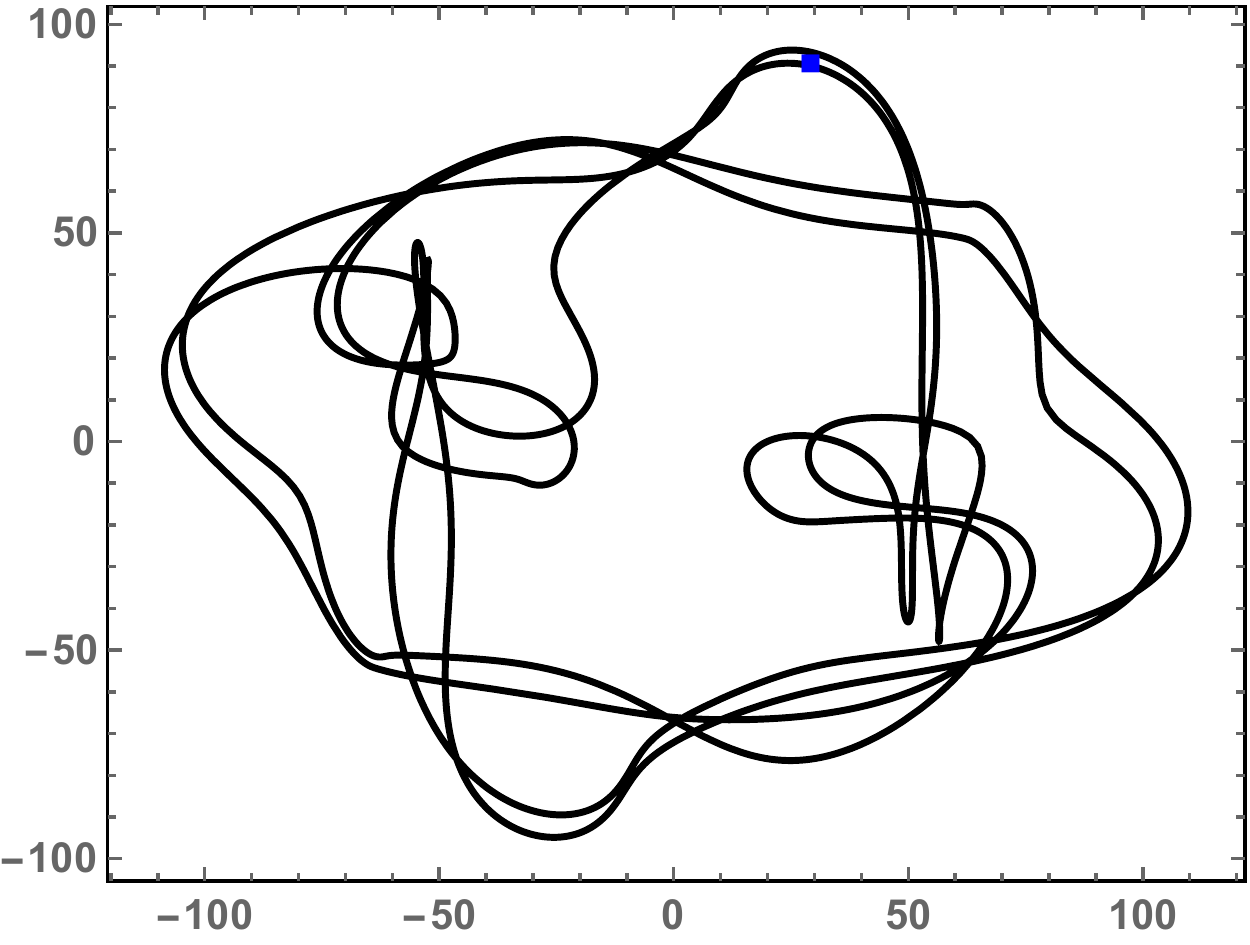}
              \caption{Initial value problem~(\ref{3Body3.2.2}),~(\ref{par:3Body3.2.2}),~(\ref{InitCond:3Body3.2.2}). Trajectory, in the complex $x$-plane, of  $x_2(t)$; 
              period $24$. The   square indicates the initial condition $x_2(0)=29.24 + 90.02  \;\mathbf{i}$.}
              \label{Ex322F4}
          \end{figure}
      \end{minipage}
  \end{minipage}
  
   \begin{minipage}{\linewidth}
      \centering
      \begin{minipage}{0.45\linewidth}
          \begin{figure}[H]
              \includegraphics[width=\linewidth]{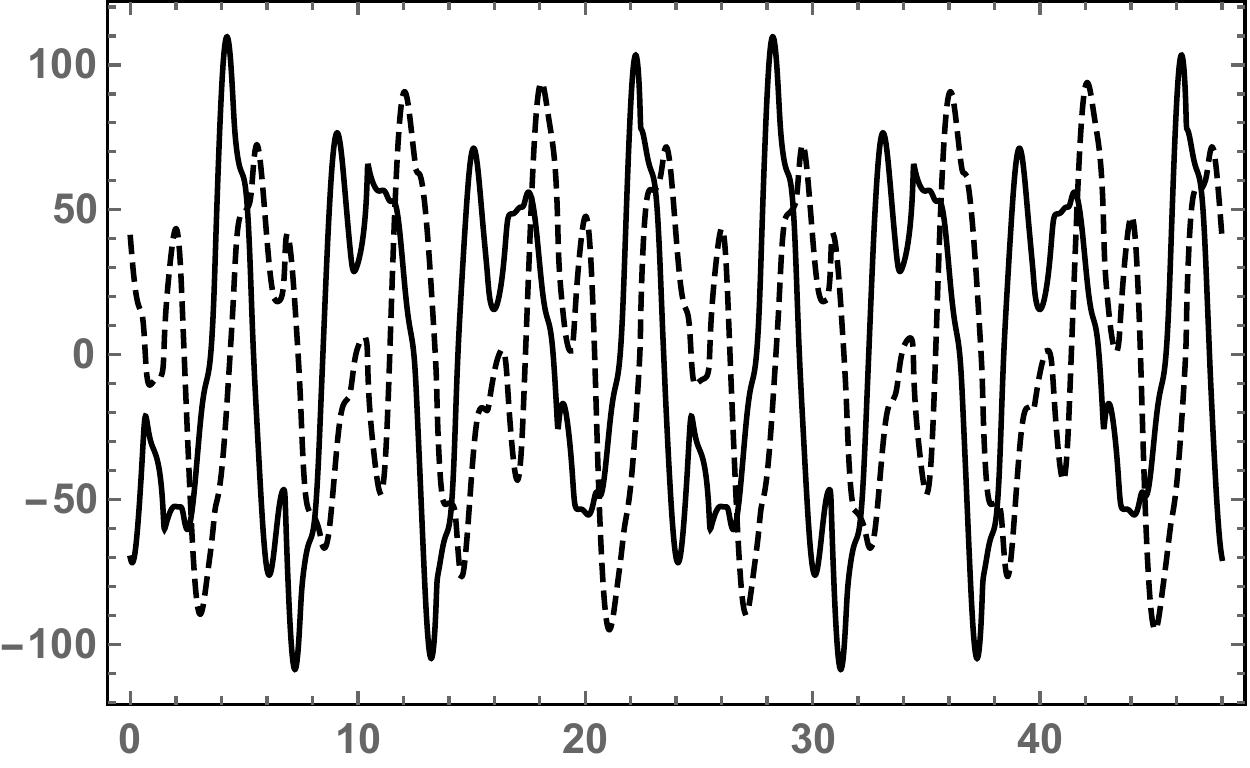}
              \caption{Initial value problem~(\ref{3Body3.2.2}),~(\ref{par:3Body3.2.2}),~(\ref{InitCond:3Body3.2.2}). Graphs of the real (bold curve) and imaginary 
              (dashed curve) parts of the coordinate $x_3(t)$; period $24$.}
              \label{Ex322F5}
          \end{figure}
      \end{minipage}
      \hspace{0.05\linewidth}
      \begin{minipage}{0.45\linewidth}
          \begin{figure}[H]
              \includegraphics[width=\linewidth]{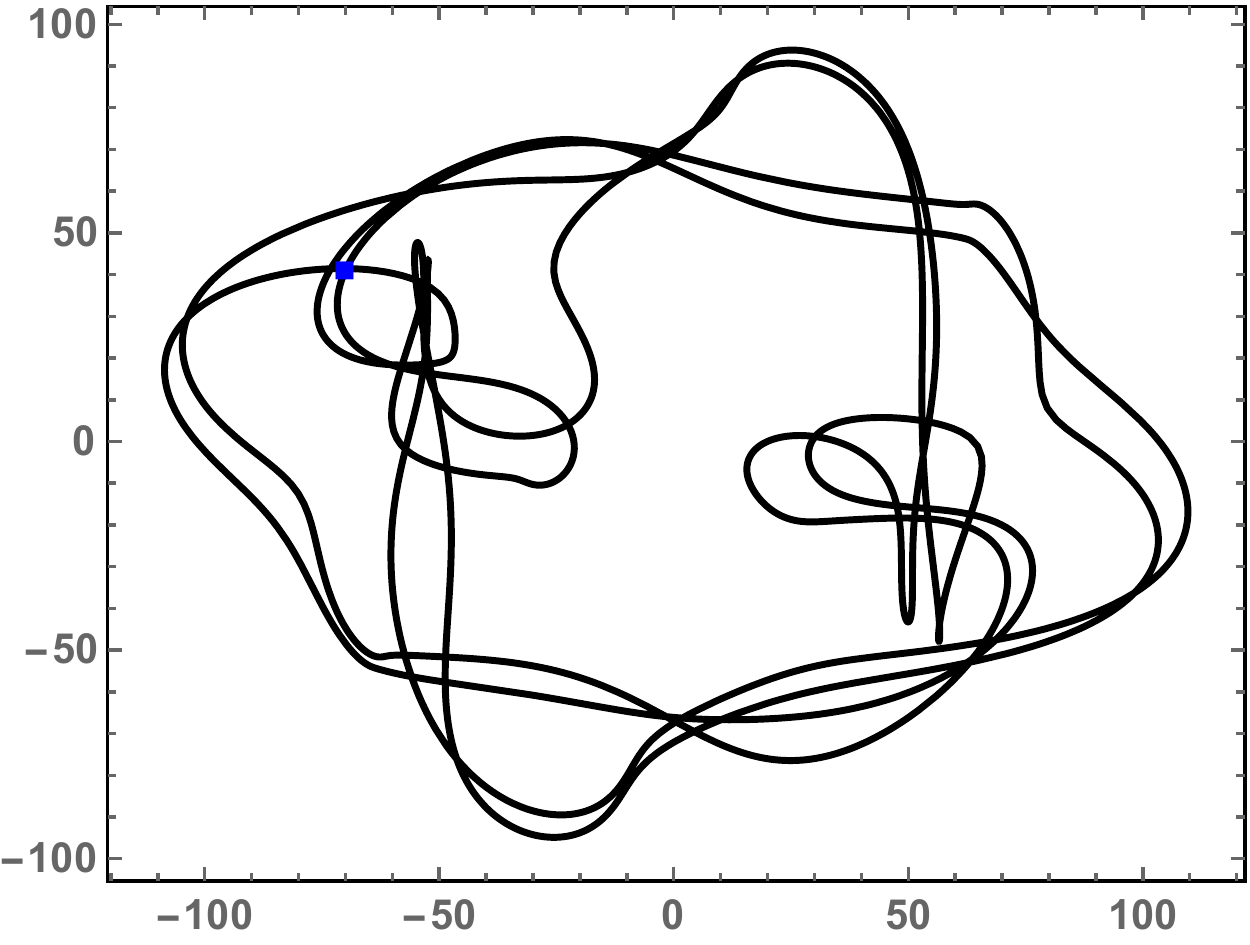}
              \caption{Initial value problem~(\ref{3Body3.2.2}),~(\ref{par:3Body3.2.2}),~(\ref{InitCond:3Body3.2.2}). Trajectory, in the complex $x$-plane, of  $x_3(t)$; 
              period $24$. The   square indicates the initial condition $x_3(0)=-70.22 + 40.41    \;\mathbf{i}$.}
              \label{Ex322F6}
          \end{figure}
      \end{minipage}
  \end{minipage}

\smallskip

\section{Discussion and Outlook}

\label{Sect4}

The results presented in this paper open several natural directions of future research.

In the present paper, for $k=1,2$, the $k$-th derivatives of the zeros $x_n(t), 1\leq n \leq N,$ of polynomial~\eqref{MainPoly} are expressed in terms of the derivatives of order $j\leq k$ of the \textit{first} $N$ coefficients $y_n, 1\leq n \leq N,$ of  polynomial~\eqref{MainPoly}, see~\eqref{xndotFormula},~\eqref{xnddotFormula}. It would be interesting to generalize these formulas for the case where $k\geq 3$ and to construct  related higher order solvable  dynamical systems.

A crucial step in obtaining formulas~\eqref{xndotFormula},~\eqref{xnddotFormula} is the solution of the overdetermined system~\eqref{Syst_yXi} for $\vec{\xi}$, by removing the last $m_1$ equations in the system as redundant. It should be possible to remove \textit{any} $m_1$ equations of system~\eqref{Syst_yXi}, to solve the resulting system  for $\vec{\xi}$ and therefore to express  $\vec{\xi}$ in terms of \textit{any} $N$ coefficients of~\eqref{MainPoly} among $y_1, \ldots, y_{N+m_1}$. Having these expressions, one may then follow the steps  outlined in Section~\ref{Sect2} to construct  first and  second order solvable dynamical systems different from those reported in Theorems~\ref{Theorem1stOrder} and~\ref{Theorem2ndOrder}.

Another natural direction is to consider, instead of~\eqref{MainPoly}, a monic time-dependent polynomial with \textit{several} multiple roots and to construct related solvable nonlinear dynamical systems.

It would be interesting  to apply a limiting procedure to known solvable dynamical systems that describe the evolution of $N$ particles $x_n(t)$ on the complex plane, to investigate the situation where two or more of the particles coalesce. 

Yet another possibility is to supplement known solvable dynamical systems  with algebraic constraints that guarantee that two or more of the particles coalesce. Dynamical systems of this kind are considered in~\cite{26,27}.

\section{Acknowledgements}

The symbolic calculations performed to obtain the main results and the examples of this paper have been verified in Mathematica. The same programming environment was used to generate  the solution plots in all the figures.

The author would also like to thank an anonymous referee of one of her previous papers for suggesting interesting directions of future research, which are included in Section~\ref{Sect4}.

\appendix

\section{The inverse of the matrix $A^{(N)}$}
\label{App1}

In this Appendix we find the inverse of the upper $N\times N$ block $A^{(N)}$ of the matrix $A$ defined by (\ref{matrixA}). Because the matrix $A^{(N)}$ is lower triangular with all its diagonal entires equal to $1$, it can be written as
\begin{equation}
A^{(N)}=I-C,
\end{equation}
where $I$ is the $N \times N$ identity matrix and the matrix $C$ is lower triangular with zero diagonal, given componentwise by
\begin{eqnarray}
&&C_{nj}=\left\{
\begin{array}{l}
\binom{m_1}{n-j} (-1)^{n+j+1} (x_1)^{n-j} \mbox{ if } j \leq n-1,\\
0 \mbox{ if } j \geq n,
\end{array}
\right.\notag\\
&& 1\leq n,j \leq N.
\end{eqnarray}
The matrix $C$ is nilpotent, indeed $C^N=0$. Therefore,
\begin{eqnarray}
\left[A^{(N)}\right]^{-1}=(I-C)^{-1}=I+\sum_{k=1}^N C^k.
\label{InverseASeries}
\end{eqnarray}
It can be shown using mathematical induction that the entries of the $k$-th power of $C$ are given by
\begin{equation}
\left[ C^k \right]_{nm}=(-1)^{n+m+1} \beta_{nm}^{(k)} (x_1)^{n-m},
\label{Ck}
\end{equation}
where the coefficients $\beta_{nm}^{(k)}$ are defined recursively by (\ref{betanm}). The formulas (\ref{InverseASeries}) and (\ref{Ck}) with (\ref{betanm}) imply (\ref{ANinverse}).

\end{document}